\documentclass[12pt]{article}
\usepackage{bm}
\usepackage{graphicx}
\usepackage{verbatim,color, array}

\usepackage{amsthm, amssymb}
\usepackage{float}
\usepackage{amsmath}
\usepackage{caption}

\usepackage{multirow}
\usepackage{multicol}

\usepackage{authblk}

\usepackage{booktabs}
\usepackage{natbib}

\usepackage{blindtext}
\usepackage{lscape}
\usepackage{pdflscape}

\newcommand{\bx}{\boldsymbol{x}}

\newcommand{\bq} {\boldsymbol{q}}

\def\boxit#1{\vbox{\hrule\hbox{\vrule\kern6pt
          \vbox{\kern6pt#1\kern6pt}\kern6pt\vrule}\hrule}}

\def\bse{\begin{eqnarray*}}
\def\ese{\end{eqnarray*}}
\def\be{\begin{eqnarray}}
\def\ee{\end{eqnarray}}
\def\bq{\begin{equation}}
\def\eq{\end{equation}}
\def\bse{\begin{eqnarray*}}
\def\ese{\end{eqnarray*}}

\begin{document}

\begin{center}
\textbf{\LARGE{
Alternative Mean Square Error Estimators and Confidence Intervals  for Prediction of Nonlinear Small Area Parameters}}
\end{center}
\vspace{0.2cm}
\center{\Large {Yanghyeon Cho and Emily Berg}}
\center{Department of Statistics, Iowa State University, 2438
Osborn Dr., Ames, IA 50011, USA}

\abstract{A difficulty in MSE estimation occurs because we do not specify a full distribution for the survey weights. This obfuscates the use of fully parametric bootstrap procedures. To overcome this challenge, we develop a novel MSE estimator. We estimate the leading term in the MSE, which is the MSE of the best predictor (constructed with the true parameters), using the same simulated samples used to construct the basic predictor. We then exploit the asymptotic normal distribution of the parameter estimators to estimate the second term in the MSE, which reflects variability in the estimated parameters. We incorporate a correction for the bias of the estimator of the leading term without the use of computationally intensive double-bootstrap procedures. We further develop calibrated prediction intervals that rely less on normal theory than standard prediction intervals. We empirically demonstrate the validity of the proposed procedures through extensive simulation studies. We apply the methods to predict several functions of sheet and rill erosion for Iowa counties using data from a complex agricultural survey. }

\section{Introduction}\label{sec1Intro}

Small area estimation refers to the practice of using model-based estimators for domains where direct estimators are considered unreliable. Many small area parameters are nonlinear functions of the response variable in the model. Important examples that occur in the domain of poverty mapping include the Gini coefficient and the proportion of the population with income below the poverty line. Nonlinear small area parameters also occur when the parameter of interest is the mean, and the model is specified in a transformed scale. For instance, a log transformation is commonly used for skewed, positive response variables.  \cite{molina2010small} propose a simulation-based procedure that approximates the empirical best predictor of general small area parameters that may be nonlinear functions of the model response variable. We call the method of \cite{molina2010small} the EBP method. \cite{molina2010small} focus on frequentist inference for the unit-level linear model. The method of \cite{molina2010small} has been extended to Bayesian inference \citep{molina2014small}, complex sampling \citep{guadarrama2018small}, two-level models \citep{marhuenda2017poverty}, generalized linear mixed models \citep{hobza2016empirical}, data-driven transformations \citep{rojas2020data}, and skew-normal models \citep{diallo2018small}. 

\cite{molina2010small} define a parametric bootstrap estimator of the mean square error (MSE) of their small area predictor.  The bootstrap MSE estimator of \cite{molina2010small} does not incorporate a correction for the bias of the estimator of the leading term, where the leading term in the MSE is the conditional variance of the small area parameter given the data. The double-bootstrap is commonly used to estimate the bias of the estimator of the leading term \citep{hall2006nonparametric, hall2006parametric}. As noted in \cite{molina2010small}, use of the double-bootstrap is often computationally prohibitive.


\hspace{.2 in } We propose an alternative way to construct MSE estimators for predictors obtained using the EBP method of \cite{molina2010small}. Our MSE estimators incorporate a correction for the bias of the estimator of the leading term, without requiring the double bootstrap. We claim that this is possible because the EBP method furnishes samples from the conditional distribution of the population parameter given the data. These samples from the conditional distribution can be used to obtain an estimator of the leading term in the MSE without implementing a bootstrap. We then use bootstrap procedures to estimate the second term in the MSE, which reflects the variation due to parameter estimation. We also use the parametric bootstrap to estimate the bias of the estimator of the leading term. The parametric bootstrap that we propose is less computationally expensive than the parametric bootstrap method of \cite{molina2010small} because we only require generating bootstrap versions of elements in the sample. In contrast, the procedure of \cite{molina2010small} requires generating a bootstrap version of the entire population. 

A further benefit of our proposed MSE estimation procedure is that it lends itself naturally to the construction of calibrated prediction intervals. \cite{molina2010small} do not consider prediction intervals explicitly. One can construct a normal-theory prediction interval as $\hat{\theta}_{i} \pm 1.96\sqrt{\hat{mse}_{i}}$, where $\hat{\theta}_{i}$ denotes the predictor and $\hat{mse}_{i}$ denotes the MSE estimator. The normal-theory prediction interval may have poor coverage if the standardized statistic defined as $T_{i} = \sqrt{\hat{mse}_{i}}^{-1}(\hat{\theta}_{i} - \theta_{i})$ does not have an approximately normal distribution, where $\theta_{i}$ denotes the true parameter.  We use the basic ingredients defining the MSE estimator to construct calibrated prediction intervals that do not require normal theory. The proposed prediction intervals adapt the calibration procedure of \cite{carlin1991sample} to the small area context. We use the same simulated samples used to estimate the leading term in the MSE to define a preliminary confidence interval. The preliminary interval is then calibrated using the bootstrap. The calibration procedure is similar to a small area prediction interval proposed in Section 2.8 of \cite{hall2006parametric}. Our procedure is tailored more specifically toward construction of intervals for nonlinear parameters under unit-level models than the method of \cite{hall2006parametric}. The prediction interval of \cite{hall2006parametric} requires a bootstrap version of the population parameter. For our procedure, we only generate bootstrap versions of sampled elements and do not construct a bootstrap version of the population parameter. Therefore, the procedure of \cite{hall2006parametric} is not tenable for use in combination with our proposed bootstrap procedure.  


A further innovation of our work is that we consider MSE estimation and confidence interval construction in the context of an informative design.  The estimator of the leading term in the MSE that we propose extends directly to an informative sample design. Estimation of the variance due to parameter estimation presents unique challenges in the context of informative sampling. In our framework, we specify only the first moment of the sample distribution of the survey weight, instead of postulating a full distribution for the survey weight. As a result, the parametric bootstrap used for the noninformative design does immediately apply in the context of informative sampling. To overcome this challenge, we simulate bootstrap parameter estimates from a nonparametric estimate of the asymptotic covariance matrix of the vector of parameter estimators. Our use of the large sample distribution of the parameter estimators allows us to circumvent the problem of specifying a full distribution for the survey weight. We also evaluate the proposed prediction intervals in the context of informative sampling. In contrast, \cite{hall2006parametric} restrict attention to noninformative designs. Extending the procedure of \cite{hall2006parametric} to informative sampling is nontrivial because the procedure of \cite{hall2006parametric} requires a bootstrap version of the population parameter.   

Variations of the proposed procedures have been used elsewhere.  \cite{sun2021bivariate} implements a version of the proposed MSE estimator in the specific context of a bivariate small area model with discrete and continuous components. Berg (2022) adapts the proposed procedure for the purpose of constructing a database for small area estimation.  The studies of \cite{sun2021bivariate} and Berg (2022) are very specific to the frameworks that they consider and are not easily generalizable. Further, \cite{sun2021bivariate} and Berg (2022) do not consider estimation of the bias of the estimator of the leading term. In this work, we generalize the procedures with the aim of reaching a broad audience. We also provide empirical and theoretical support for the methodology. We conduct a thorough empirical evaluation of several estimators of the bias of the estimator of the leading term in the MSE. 

Upon completing this work, we learned that the estimator of the leading term that we propose is in current use for production of poverty indicators at the World Bank. The World Bank MSE estimator, however, does not appropriately reflect variability due to parameter estimation. One of the contributions of our study is to provide rigorous support for the estimator of the leading term in the MSE that is currently in use at the World Bank (Isabelle Molina, Personal Communication, 7-6-22). The estimator of the second term in the MSE that we propose has potential use of inference about poverty measures at the World Bank. The relevance of the proposed procedures to the current practice at the World Bank demonstrates that the methodology in this paper is of salient importance for statistical practice. 

Many other works propose MSE estimators that incorporate corrections for the bias of the estimator of the leading term. \cite{hall2006nonparametric} and \cite{hall2006parametric} propose parametric and non-parametric double-bootstrap based MSE estimators that are very computationally intensive to implement. We do not consider the bootstrap procedures of \cite{hall2006nonparametric} or \cite{hall2006parametric} as a result of the computational burden. To reduce the computational demands, \cite{erciulescu2016small} develop a fast double bootstrap. The fast double bootstrap MSE estimator of \cite{erciulescu2016small} can result in negative estimates.  In a study of small area estimation based on the gamma distribution, Cho and Berg (in prepration) find that the prevalence of negative estimates from the method of \cite{erciulescu2016small} is nontrivial. Because one cannot construct a confidence interval from a negative MSE estimate, we do not consider the method of \cite{erciulescu2016small}. \cite{erciulescu2019bootstrap} develop calibrated confidence intervals for small area means. It is not immediately obvious to us that the method of \cite{erciulescu2019bootstrap} extends to nonlinear small area parameters. \cite{lahiri2007resampling} develop positive MSE estimates that incorporate a correction for the bias of the estimator of the leading term in the context of an area-level model. An extension of their method to prediction of nonlinear parameters in the context of a unit-level model is not straightforward and is beyond the scope of our work. Further, their MSE estimator is much more difficult to implement than the MSE estimator that we propose.  An alternative to the bootstrap is to use the jackknife to estimate the bias of the estimator of the leading term \citep{lohr2009jackknife}.  The jackknife MSE estimator of \cite{lohr2009jackknife} is developed for an area-level model. Because we focus on unit-level models, we do not consider the jackknife MSE estimator of \cite{lohr2009jackknife}. The SUMCA method \citep{jiang2020sumca} is an alternative way to construct a bias correction. We do not consider the SUMCA method because it is not clear to us that SUMCA appropriately reflects the variance of parameter estimators for nonlinear parameters, as we explain in Appendix A of the supplementary material (SM). 

We propose inference procedures that are computationally simple to implement when used in  combination with the EBP procedure of \cite{molina2010small}. The procedures lend themselves naturally to construction of calibrated prediction intervals and informative sampling.  In Section 2, we define the proposed method for non-informative and informative sample designs. We also define the confidence intervals and corrections to the bias of the estimator of the leading term in Section 2.  In Section 3, we evaluate the proposed procedure through simulations that use both noninformative and informative designs. In Section 4, we present two data analyses: one for non-informative sampling and a second for informative sampling.  

\section{Proposed Method}\label{sec2Proposed}

As a precursor to defining the proposed MSE estimator and confidence interval, we overview the method of \cite{molina2010small} in Section 2.1. The development of Section 2.1 is slightly more general than the development in \cite{molina2010small}. We define the procedure for a general model, while \cite{molina2010small} focus on the unit-level linear model. Nonetheless, the basic concepts used in Section 2.1 are essentially the same as those in \cite{molina2010small}. After we overview the basic approach of \cite{molina2010small}, we define the proposed MSE estimator and confidence interval in Section \ref{sec2.2ProposedMSECI}, \ref{sec2.4BC}, and \ref{sec2.5CI}. We extend the procedures to an informative sample design in Section \ref{sec2.3Informative}.  

\subsection{ Overview of EBP Method of Molina and Rao (2010)}\label{sec2.1Molina}

\hspace{.2 in } Let $i = 1,\ldots, D$ index small areas, and let $j = 1,\ldots, N_{i}$ index the population elements in small area $i$. Let $y_{ij}$ be the variable of interest, and define the small area parameter $\theta_{i}$ by  $\theta_{i} = h(y_{i 1},\ldots, y_{i N_{i}})$. Without loss of generality, assume that the first $n_{i}$ elements are sampled, and denote the sampled elements by $\bm{y}_{si} = (y_{i1},\ldots, y_{i n_{i}})'$.  Let a covariate $\bm{x}_{ij}$ be known for the full population of $N_{i}$ elements in area $i$. Assume that $y_{ij}$ satisfies the model  $y_{ij}\sim f(y_{ij} \mid b_{i}, \bm{x}_{ij}; \bm{\psi}_{1}),$ where $f(\cdot \mid \cdot)$ is an appropriately specified pdf/pmf, and $b_{i}\sim N(0,\sigma_{b}^{2})$. Assume a sample is selected from area $i$, and let $j = 1,\ldots, n_{i}$ index the sampled elements in area $i$.  Assume that an estimator $\hat{\bm{\psi}} = (\hat{\bm{\psi}}_{1}', \hat{\sigma}^{2}_{b})'$ of $\bm{\psi} = (\bm{\psi}_{1}', \sigma^{2}_{b})'$ is constructed with the sampled data using a method such as maximum likelihood. The estimator of $\theta_{i}$ given in \cite{molina2010small} is defined as $\hat{\theta}_{i}^{EBP} = L^{-1}\sum_{\ell = 1}^{L} \hat{\theta}_{i}^{(\ell)}$, where $\hat{\theta}_{i}^{(\ell)}\sim f(\theta_{i} \mid \bm{y}_{si}, \bm{x}_{N_{i}}, \hat{\bm{\psi}})$, $\bm{y}_{si} = (y_{i1},\ldots, y_{i n_{i}})'$, and $\bm{x}_{N_{i}} = \{\bm{x}_{ij}: j=1,\hdots, N_{i}\}$. At this stage, \cite{molina2010small} turn to the parametric bootstrap for uncertainty estimation. We propose an alternative MSE estimator and confidence interval. 



 
\subsection{  Proposed MSE Estimator and Confidence Interval}\label{sec2.2ProposedMSECI}

\hspace{.2 in } {\it Theorem 1} gives a decomposition of the MSE of the EBP predictor that will be central to our development of the MSE estimator. 

{\it Theorem 1}: The MSE of the EBP predictor can be expressed as
\begin{align}\label{msedecomp1}
{\rm MSE}(\hat{\theta}_{i}^{EBP}) &= E[ V\{\theta_{i} \mid \bm{y}_{si}; \bm{x}_{N_{i}}, {\bm{\psi}}\}] + E[(\hat{\theta}_{i}^{EB} - \hat{\theta}_{i}^{B})^{2}] + E[(\hat{\theta}_{i}^{EBP} - \hat{\theta}_{i}^{EB})^{2}] , \\ \nonumber 
                & =  M_{1i} +  M_{2i} + M_{3i},
\end{align}
where  $\hat{\theta}_{i}^{B}  = E[\theta_{i} \mid \bm{y}_{si}; \bm{x}_{N_{i}}, \bm{\psi}],$ 
$E[(\hat{\theta}_{i}^{EBP} - \hat{\theta}_{i}^{EB})^{2}] = L^{-1}E[V\{\theta_{i} \mid \bm{y}_{si}; \bm{x}_{N_{i}}, \hat{\bm{\psi}}\}],$ and $\hat{\theta}_{i}^{EB} = E[\theta_{i} \mid \bm{y}_{si}; \bm{x}_{N_{i}}, \hat{\bm{\psi}}]$. 

\qedsymbol 

A proof of Theorem 1 is given in Appendix B of the SM. We call the first term in the MSE ($M_{1i}$) the leading term. This term is typically the dominant term in the MSE and is the mean square error of the optimal predictor, $\hat{\theta}_{i}^{B}$.  The second term ($M_{2i}$) reflects the increase in MSE due to replacement of $\bm{\psi}$ with $\hat{\bm{\psi}}$. We ignore the last term in the MSE ($M_{3i}$) under the presumption that $L$ can be taken to be arbitrarily large.  

We estimate the leading term and the second term separately. We then construct an estimate of the bias of the estimator of the leading term. The approach of estimating the first two terms in the MSE separately is related to bootstrap MSE estimators proposed in \cite{butar2003measures}, \cite{gonzalez2007estimation}, \cite{lahiri2007resampling}, and \cite{booth1998standard}. 

We first construct an estimator of the leading term. Our innovation is to use the EBP concept to estimate the first term in the MSE. This allows us to estimate the bias of the estimator of the leading term, without requiring the double bootstrap.  The crux of the estimator of the leading term is the observation that $\hat{\theta}_{i}^{(\ell)}$ for $\ell = 1,\ldots, L$ are $iid$, $E_{L}[\hat{\theta}_{i}^{(\ell)} \mid \bm{y}_{si}, \bm{x}_{N_{i}}, \hat{\bm{\psi}}] = E[ \theta_{i}  \mid \bm{y}_{si}, \bm{x}_{N_{i}}, \hat{\bm{\psi}}]$ and $V_{L}(\hat{\theta}_{i}^{(\ell)}) = V(\theta_{i} \mid \bm{y}_{si}, \bm{x}_{N_{i}}, \hat{\bm{\psi}})$,  where $E_{L}$ and $V_{L}$ denote expectation and variance relative to the distribution used to generate $\hat{\theta}_{i}^{(\ell)}$ for $\ell = 1,\ldots, L$. By standard properties of $iid$ random variables, the sample variance of $\hat{\theta}_{i}^{(\ell)}$ for $\ell = 1,\ldots, L$ is an unbiased estimator of $V\{\theta_{i} \mid \bm{y}_{si}, \bm{x}_{N_{i}}, \hat{\bm{\psi}}\}$. This justifies an estimator of the leading term in the MSE defined as  
\begin{align}\label{leadingtermest}
\hat{M}_{1i}  = \frac{1}{L-1}\sum_{\ell= 1}^{L}(\hat{\theta}_{i}^{(\ell)} - \hat{\theta}_{i}^{EBP})^{2}.
\end{align}
{\it Theorem 2 } formalizes the properties of the estimator of $M_{1i}$. The proof of {\it Theorem 2} is given in Appendix C of the SM. 

{\it Theorem 2: } Assume $\hat{\bm{\psi}} \stackrel{p}{\rightarrow} \bm{\psi}$ as $D\rightarrow\infty$.  Assume $V\{\theta_{i} \mid \bm{y}_{si}, \bm{x}_{N_{i}}, \bm{\psi}\}$ is a differentiable function of $\bm{\psi}$. Assume 
\begin{align*}
E[(D_{1}(\bm{\psi}^{*}, \bm{y}_{si})(\hat{\bm{\psi}} - \bm{\psi}) )^{2}] = o(1), 
\end{align*}
where $\bm{\psi}^{*}$ is in a closed ball containing $\bm{\psi}$, and $D_{1}(\bm{\psi}, \bm{y}_{si}) = \partial V\{\theta_{i} \mid \bm{y}_{si}, \bm{x}_{N_{i}}, \bm{\psi}\}/\partial \bm{\psi} $. Assume the function $h$ is such that $E_{L}[\hat{\theta}_{i}^{(\ell)} \mid \bm{y}_{si}, \bm{x}_{N_{i}}, \hat{\bm{\psi}}] <\infty$ and $V_{L}(\hat{\theta}_{i}^{(\ell)}) < \infty$.  Then, we have the following two results:
\begin{enumerate}
\item[(i)] $E_{L}[\hat{M}_{1i} -V\{\theta_{i} \mid \bm{y}_{si}, \bm{x}_{N_{i}}, \hat{\bm{\psi}}\} \mid \bm{y}_{si}, \bm{x}_{N_{i}}, \hat{\bm{\psi}}] = 0$.
\item[(ii)] $|\hat{M}_{1i}  - V\{\theta_{i} \mid \bm{y}_{si}, \bm{x}_{N_{i}}, \hat{\bm{\psi}}\}|\stackrel{p}{\rightarrow} 0$ as $D \rightarrow\infty$. 
\end{enumerate}

\qedsymbol

We next define an estimator of $M_{2i}$. We use a modification of a parametric bootstrap procedure to account for variability in $\hat{\bm{\psi}}$. The parametric bootstrap procedure involves implementing the following steps for $b = 1,\ldots, B$:
\begin{enumerate}
\item First, generate a bootstrap sample. Simulate $b_{i}^{(b)}\sim N(0,\hat{\sigma}^{2}_{b})$, and generate $y_{ij}^{*(b)}\sim f(y_{ij} \mid b_{i}^{(b)}, \bm{x}_{ij}, \hat{\bm{\psi}})$ for $j = 1,\ldots, n_{i}$. 
\item Repeat the estimation procedure (i.e., maximum likelihood) using $\{y_{ij}^{*(b)}: j = 1,\ldots, n_{i}\}$ to obtain bootstrap parameter estimates $\hat{\bm{\psi}}^{(b)}$. 
\item Re-compute predictors and estimators of the leading term in the MSE using the bootstrap parameters and the original data, $\bm{y}_{si}$. Specifically, define a bootstrap version of the predictor by  $\hat{\theta}_{i}^{ (b)} = L^{-1}\sum_{\ell = 1}^{L} \hat{\theta}_{i}^{(\ell, b)}$, where $\hat{\theta}_{i}^{(\ell, b)} \sim f(\theta_{i} \mid \bm{y}_{si}, \bm{x}_{N_{i}}, \hat{\bm{\theta}}^{(b)})$. Likewise, define a bootstrap version of the estimator of the leading term in the MSE by $\hat{M}_{1i}^{(b)} = (L-1)^{-1}\sum_{\ell = 1}^{L}(  \hat{\theta}_{i}^{(\ell, b)} - \hat{\theta}_{i}^{(b)})^{2}$.  
\end{enumerate}
Define a bootstrap estimator of the second term in the MSE  by 
\begin{align}\label{Mhat2i}
\hat{M}_{2i,L} = B^{-1}\sum_{b= 1}^{B}(\hat{\theta}_{i}^{(b)} - \hat{\theta}_{i})^{2}, 
\end{align}
where the subscript of $L$ is used to indicate that the estimator of the second term in the MSE depends implicitly on the $L$ simulated samples. 

Note that in step 1 of the parametric bootstrap procedure, we only generate $y_{ij}$ for sampled elements and not for elements of the full population. In contrast, the parametric bootstrap procedure of \cite{molina2010small} requires generating $y_{ij}^{(b)}$ for all elements in the population. Therefore, our parametric bootstrap method is less computationally demanding than that of \cite{molina2010small}. 

In Step 3 of the procedure, we use the bootstrap maximum likelihood estimator but the original data. We do this for two reasons. First, the estimator of the second term in the MSE accounts for the variation of the parameter estimators. To isolate the effect of the variance of the parameter estimators, we hold the original data fixed. Second, the use of the original data permits us to obtain a computationally efficient estimator of the second term in the MSE. We only need to calculate the empirical best predictor once for each boostrap sample, and we do not need to generate a bootstrap version of the entire population.  

Theorem 3 provides further insight into why we use the bootstrap parameter estimates, while holding the original data fixed when defining the estimator of the second term in the MSE. In Theorem 3, we prove that the bootstrap MSE estimator of the second term in the MSE is asymptotically equivalent to a Taylor approximation variance estimator. For simplicity, we develop Theorem 3 for the extreme case in which $L = \infty$.  To state Theorem 3, we require the definitions of the following quantities. Let $\hat{\theta}_{i,\infty}^{(b)} = E[\theta_{i} \mid \bm{y}_{si}, \bm{x}_{N_{i}}, \hat{\bm{\psi}}^{(b)} ]$. Define $\hat{g}_{2i,\infty} = B^{-1}\sum_{ b = 1}^{B}(\hat{\theta}_{i,\infty}^{(b)} - \hat{\theta}_{i}^{EB}  )^{2}$, where $\hat{\theta}_{i}^{EB}$ is defined in the statement of {\it Theorem 1}. The proof of Theorem 3 is given in Appendix D of the SM.

{\it Theorem 3}:    For given $\bm{\psi}$, let $g(\bm{\psi}, \bm{y}_{si}, \bm{x}_{N_{i}}) = E[\theta_{i} \mid \bm{y}_{si}, \bm{x}_{N_{i}}, \bm{\psi}]$. Let $\dot{\bm{g}}(\bm{\psi}, \bm{y}_{si}, \bm{x}_{N_{i}}) = \partial g(\bm{\psi},\bm{y}_{si}, \bm{x}_{N_{i}})/(\partial \bm{\psi})$. Assume that the regularity conditions in the supplement hold. Then, 
\begin{align*}
    \hat{M}_{2i,\infty} =  \dot{\bm{g}}(\hat{\bm{\psi}},\bm{y}_{si}, \bm{x}_{N_{i}})'V\{\hat{\bm{\psi}}\}\dot{\bm{g}}(\hat{\bm{\psi}},\bm{y}_{si}, \bm{x}_{N_{i}}) + o_{p}(D^{-1}),
\end{align*}
 where $\hat{M}_{2i,\infty} = \lim_{L\rightarrow\infty}\hat{M}_{2i,L}$. 

\qedsymbol 

We combine the estimator of $M_{1i}$ with the estimator of $M_{2i}$ to define an MSE estimator. Define an MSE estimator by 
\begin{align}\label{MSEhatNoBC} 
    \widehat{\rm MSE}_{i}^{\rm no\_BC} = \hat{M}_{1i} + \hat{M}_{2i,L},  
\end{align}
where $\hat{M}_{1i}$ is defined in (\ref{leadingtermest}) and $\hat{M}_{2i,L}$ is defined in (\ref{Mhat2i}). The superscript ${\rm no\_BC}$ is used to differentiate (\ref{MSEhatNoBC}) from MSE estimators proposed below in subsection 2.4 that apply bias corrections to the estimator of the leading term.

\subsection{ Extension to Informative Sampling}\label{sec2.3Informative}

The proposed  MSE estimator and confidence interval extend readily to an informative sample design. Assume $d < D$ areas is selected, and assume $n_{i}$ elements is selected from each sampled area.   Let $I_{i}$ and $I_{ij}$ be sample inclusion indicators for areas and units within areas, respectively.  Let $s = \{i: I_{i} = 1\}$ and let $s_{i} = \{j: I_{ij} = 1\}$. Assume that a model for the sample distribution is specified as $y_{ij} \mid    (I_{ij} = 1 ) \sim f_{sy}(y_{ij} \mid \bm{x}_{ij}, b_{i}, I_{ij} = 1, \bm{\psi}_{s1})$, where $b_{i} \mid (I_{i} = 1)\sim N(0,\sigma^{2}_{b})$.  Let $w_{ij}^{-1} = \pi_{ij}^{-1}$ and $w_{i}^{-1} = \pi_{i}^{-1}$, where $\pi_{ij} = P(I_{ij} = 1 \mid \bm{x}_{ij}, y_{ij})$ and $\pi_{i} = P(I_{i} = 1 \mid b_{i})$. Assume  $E[w_{ij} \mid (I_{ij} = 1)] = \omega ( y_{ij}, \bm{x}_{ij}, \bm{\psi}_{s2})$. For the purpose of prediction for nonsampled areas, we postulate a lognormal model for $w_{i}$  and assume $\mbox{log}(w_{i}) \sim N(\lambda_{0} + \lambda_{1}b_{i}, \tau^{2})$. Let $\bm{\psi}_{3} = (\lambda_{0}, \lambda_{1}, \tau^{2})'$. 

The prediction procedure requires estimators of $\bm{\psi}= (\bm{\psi}_{s1}',\bm{\psi}_{s2}', \bm{\psi}_{s3}', \sigma^{2}_{b} )'$. Let $(\hat{\bm{\psi}}_{s1}', \hat{\sigma}^{2}_{b})$ be the maximum likelihood estimator defined by 
\begin{align*}
(\hat{\bm{\psi}}_{s1}', \hat{\sigma}^{2}_{b})' = argmax_{(\bm{\psi}_{s1}', \sigma^{2}_{b})} \prod_{i \in s}\int_{\infty}^{\infty} \prod_{j \in s_{i}} 
f_{sy}(y_{ij} \mid \bm{x}_{ij}, b_{i}, I_{ij} = 1, \bm{\psi}_{s1})\phi(b_{i}/\sigma_{b})/\sigma_{b}d b_{i},
\end{align*}
where $\phi(\cdot)$ is the pdf of a standard normal random variable.  Define the estimator of $\bm{\psi}_{s2}$ as $\hat{\bm{\psi}}_{s2} = argmin_{\bm{\psi}_{s2}} \sum_{i\in s}\sum_{j\in s_{i}} (w_{ij} - \omega ( y_{ij}, \bm{x}_{ij}, \bm{\psi}_{s2}))^{2}$. Estimate $\bm{\psi}_{s3}$ by 
\begin{align*}
\hat{\bm{\psi}}_{s3} = argmax_{\bm{\psi}_{s3} } \prod_{i\in s} \int_{-\infty}^{\infty}\frac{1}{\tau} \phi\left( \frac{ \mbox{log}(w_{i}) - \lambda_{0} - \lambda_{1}b_{i}}{\tau}  \right)\prod_{j\in s_{i}} f_{sy}(y_{ij} \mid \bm{x}_{ij}, b_{i}, I_{ij} = 1, \hat{\bm{\psi}}_{s1})\phi(b_{i}/\hat{\sigma}_{b})/\hat{\sigma}_{b}db_{i}.
\end{align*}

To construct the predictors, we require samples from the distribution nonsampled elements given the observed data.  Let $\bm{y}_{ci} = \{y_{ij}: I_{ij} = 0, I_{i} = 1\}$ and $\bm{y}_{N_{i}} = \{y_{ij}: j = 1,\ldots, N_{i}\}$. Let $D_{s}$ denote the set of observed data, as in \cite{pfeffermann2007small}. Let $\bm{y}_{si} = \{y_{ij}: I_{ij} = 1, I_{i} = 1\}$.   By a generalization of \cite{pfeffermann2007small} (see Cho et al., undated), one can express the required conditional distributions as functions of the sample distributions. The required distribution for sampled areas is of the form 
\begin{align}\label{compsamp1} 
    f_{p}( y_{ij} \mid D_{s}, I_{i} = 1, I_{ij} = 0, \bm{\psi}) =  \int_{-\infty}^{\infty} \frac{(\omega(y_{ij}, \bm{x}_{ij}, \bm{\psi}_{s2}) - 1)  f_{sy}(y_{ij} \mid \bm{x}_{ij}, b, I_{ij} = 1)}{(\omega(y_{ij}, \bm{x}_{ij}, \bm{\psi}_{s2}) - 1)   } f_{si}(b  \mid D_{s}, \bm{\psi}_{s1}, \sigma^{2}_{b}) db, 
\end{align}
where $y_{ij} \perp y_{ik} \mid (D_{s}, I_{i} = 1, I_{ij} = 0, \bm{\psi})$ for $(j,k)\notin s_{i}$, and 
\begin{align}\label{bcond}
f_{si}(b  \mid D_{s}, \bm{\psi}_{s1}, \sigma^{2}_{b})  = \frac{ [\prod_{j \in s_{i} } f_{sy}(y_{ij} \mid \bm{x}_{ij}, b, I_{ij} = 1)  ]\phi(b/\sigma_{b})/\sigma_{b}}{  \int_{-\infty}^{\infty}  [\prod_{j \in s_{i} } f_{sy}(y_{ij} \mid \bm{x}_{ij}, b, I_{ij} = 1)  ]\phi(b/\sigma_{b})/\sigma_{b}d b}. 
\end{align}
 The required distribution for nonsampled areas is of the form 
\begin{align}\label{compsamp2} 
    f_{p}(y_{ij} \mid D_{s}, I_{i} = 0, \bm{\psi}) =  \int_{-\infty}^{\infty} \frac{ E_{s}[\pi_{ij}^{-1}  \mid y_{ij}, \bm{x}_{ij}, b_{i},I_{i}=1] f_{si}(y_{ij} \mid \bm{x}_{ij}, u_{i},I_{i}=1)}{E_{si}[\pi_{ij}^{-1} \mid \bm{x}_{ij}, u_{i},I_{i}=1]  }\frac{ E_{s}[\pi_{i}^{-1}-1 \mid u_{i}]}{ E_{s}[\pi_{i}^{-1} - 1]  }f_{s}(u_{i}) d u_{i},   
\end{align}
where $y_{ij} \perp y_{ik} \mid (D_{s}, I_{i} = 0, \bm{\psi}) $.  Assume a procedure is available for sampling from an estimate of the distributions (\ref{compsamp1}) and (\ref{compsamp2}). Cho et al. (in preparation) define a general procedure that uses sampling importance resampling \citep{smith1992bayesian}. For a sampled area, define $\theta_{i}^{(\ell)} =  h(y_{i1}^{(\ell)},\ldots, y_{i N_{i}}^{(\ell)})$, where $y_{ij}^{(\ell)} = 1$ if $I_{ij} = 1$ and $y_{ij}^{(\ell)} \sim  f_{p}( y_{ij} \mid D_{s}, I_{i} = 1, I_{ij} = 0, \bm{\psi})$ if $I_{ij} = 0$. For a non-sampled area, generate $y_{ij}^{(\ell)} \sim  f_{p}(y_{ij} \mid D_{s}, I_{i} = 0, \bm{\psi})$ for $j = 1,\ldots, N_{i}$. Then, an EBP of $\theta_{i}$ for the informative design is given by $\hat{\theta}_{i}  = L^{-1}\sum_{\ell = 1}^{L} \hat{\theta}_{i}^{(\ell)}$.

The simulation procedure immediately furnishes an estimate of the leading term in the MSE defined as $\hat{M}_{1i} = (L-1)^{-1}\sum_{\ell = 1}^{L}(\hat{\theta}_{i}^{(\ell)} - \hat{\theta}_{i})^{2}.$ The properties of the estimator of the leading term given in {\it Theorem 2} apply directly to the informative design. The proof of Theorem 2 carries over to the informative design by replacing $(\bm{y}_{si}, \bm{x}_{N_{i}})$ with $(D_{s}, I_{i})$. 

A challenge in constructing the second term in the MSE occurs because we do not have a full distribution for $w_{ij}$. This complicates the problem of defining a bootstrap version of the weight. We overcome this issue by implementing a slight modification to the estimator of the second term in the MSE.  Decompose $\bm{\psi}$ as $\bm{\psi} = (\bm{\psi}_{\infty}', \bm{\psi}_{pos}')'$, where $\bm{\psi}_{\infty}$ denotes the components of $\bm{\psi}$ with parameter space $(-\infty, \infty)$, and $\bm{\psi}_{pos}$ denotes the components of $\bm{\psi}$ with parameter space $(0,\infty)$. (We assume that all components of $\bm{\psi}$ have parameter space either $(-\infty,\infty)$ or $(0, \infty)$.) For a vector $\bm{v} = (v_{1}, \ldots, v_{p})'$ with positive components, let $\mbox{log}(\bm{v}) = (\mbox{log}(v_{1}), \ldots, \mbox{log}(v_{p}))'$ and let $\mbox{exp}(\mbox{log}(\bm{v})) = (v_{1}, \ldots, v_{p})'$. Let $\hat{\bm{\psi}}_{T} = (\hat{\bm{\psi}}_{\infty}', log(\hat{\bm{\psi}}_{pos})')'$ and let $\hat{V}(\hat{\bm{\psi}}_{T})$ denote an estimate of the variance of the asymptotic normal distribution of $\hat{\bm{\psi}}_{T}$. In the simulations and data analysis, we use the jackknife to obtain $\hat{V}(\hat{\bm{\psi}}_{T})$. For bootstrap samples $b = 1,\ldots, B$, simulate $\hat{\bm{\psi}}_{T}^{(b)} = ( (\hat{\bm{\psi}}_{\infty}^{(b)})', \mbox{log}(\hat{\bm{\psi}}_{pos}^{(b)})')  \stackrel{iid}{\sim} N(\hat{\bm{ \psi}}_{T}, \hat{V}(\hat{\bm{\psi}}_{T}) )$. Set $\hat{\bm{\psi}}^{(b)} = ( (\hat{\bm{\psi}}_{\infty}^{(b)})',  \mbox{exp}(\mbox{log}(\hat{\bm{\psi}}_{pos}^{(b)}))')$. For sampled areas, generate $\hat{\theta}_{i}^{(\ell, b)} =  h(y_{i1}^{(\ell,b)},\ldots, y_{iN_{i}}^{(\ell,b)})$, where $y_{ij}^{(\ell, b)} \sim f_{p}(y_{ij} \mid D_{s}, I_{i} = 1, I_{ij} = 0, \hat{\bm{\psi}}^{(b)})$ for $j \notin s_{i}$ and $y_{ij}^{(\ell, b)} = y_{ij}^{(b)} for j\in s_{i}$.  For nonsampled areas, generate $\hat{\theta}_{i}^{(\ell, b)} =  h(y_{i1}^{(\ell,b)},\ldots, y_{iN_{i}}^{(\ell,b)})$, where $y_{ij}^{(\ell, b)} \sim f_{p}(y_{ij} \mid D_{s}, I_{i} = 0, \hat{\bm{\psi}})$ for $j= 1,\ldots, N_{i}$. Let $\hat{\theta}_{i}^{(b)} = L^{-1}\sum_{\ell = 1}^{L}\hat{\theta}_{i}^{(\ell,b)}$. An estimate of the second term in the MSE is $\hat{M}_{2i,L} =  B^{-1}\sum_{b=1}^{B}(\hat{\theta}_{i}^{(b)}  - \hat{\theta}_{i})^{2}.$.  An estimator of the MSE that does not incorporate a correction for the bias of the estimator of the leading term is given by 
\begin{align}\label{nobcInf}
\widehat{\rm MSE}_{i}^{\rm no\_BC, Inf} = \hat{M}_{1i} + \hat{M}_{2i,L}. 
\end{align}
Define an estimate of the leading term in the MSE for replicate $b$ by 
\begin{align}\label{Mhat1ibInf}
\hat{M}_{1i}^{(b)} = (L-1)^{-1}\sum_{\ell = 1}^{L}(\hat{\theta}_{i}^{(\ell,b)} - \hat{\theta}_{i}^{(b)})^{2}.
\end{align}
We use $\hat{M}_{1i}^{(b)}$ in the next subsection to construct estimates of the bias of the estimator of the leading term. 

\subsection{ Bias Corrections for $\hat{M}_{1i}$}\label{sec2.4BC}

\hspace{0.2 in} A problem with $\widehat{\rm MSE}_{i}^{\rm no\_BC}$ is that $E[\hat{M}_{1i} - M_{1i}] \neq 0$. We need to consider the bias induced by the replacement of ${\bm{\psi}}$ with $\hat{\bm{\psi}}$ when estimating the leading term in the MSE. We define several bias-corrected estimators for $\hat{M}_{1i}$ that use different types of bias adjustments.  The bias corrections use the bootstrap versions of the estimates of the leading term in the MSE defined as $\hat{M}_{1i}^{(b)}$, where $\hat{M}_{1i}^{(b)}$ is defined in step 3 of the bootstrap procedure for the noninformative design and $\hat{M}_{1i}^{(b)}$ is defined in (\ref{Mhat1ibInf}) for the informative design.  

An additive bias-corrected estimator is $   \hat{M}_{1i}^{Add} = \hat{M}_{1i}-\big\{\bar{M}_{1i}^{*}-\hat{M}_{1i}\big\} =2\hat{M}_{1i}-\bar{M}_{1i}^{*},$ where $\bar{M}_{1i}^{*} = B^{-1}\sum_{b=1}^{B}\hat{M}_{1i}^{(b)}$. This customary bias correction adjusts the bias by adding the estimate of the bias, $E\big[\hat{M}_{1i}-M_{1i}\big]$. A possible problem with the additive adjustment is that the resulting MSE estimator can be negative. To avoid negative estimates, we define a multiplicative bias-corrected estimator by  $\hat{M}_{1i}^{Mult} =  \hat{M}_{1i}^{2}(\bar{M}_{1i}^{*})^{-1}$.  This estimator adjusts the bias through multiplication by an estimate of the factor, $E\big[\hat{M}_{1i}\big]/M_{1i}.$ The multiplicative correction may be unstable if the denominator of the adjustment factor is close to zero.  This leads us to  define a compromise between the multiplicative and additive adjustments by 
\begin{align}\label{compdef} 
 \hat{M}_{1i}^{Comp} &=
 \begin{cases}
 \hat{M}_{1i}^{Add}, & \text{if } \hat{M}_{1i} \geq \bar{M}_{1i}^{*},\\
 \hat{M}_{1i}^{Mult}, & \text{if }  \hat{M}_{1i} < \bar{M}_{1i}^{*}.
 \end{cases}
\end{align}
We expect that this compromise between the multiplicative and additive corrections will avoid negative or infinity MSE estimates. The form of the compromise bias correction (\ref{compdef}) is a specific type of the general class of bias corrections proposed in \cite{hall2006nonparametric}.  Lastly, we borrow a different form for the bias correction from \cite{hall2006parametric}. We denote it as 
\begin{align*}
 \hat{M}_{1i}^{HM}(\hat{\bm{\psi}}) &=
 \begin{cases}
 2\hat{M}_{1i}-\bar{M}_{1i}^{*}, & \text{if } \hat{M}_{1i} \geq \bar{M}_{1i}^{*},\\
 \hat{M}_{1i} exp\big[-\big\{\bar{M}_{1i}^{*}-\hat{M}_{1i}\big\}/\bar{M}_{1i}^{*}\big], & \text{if }  \hat{M}_{1i} < \bar{M}_{1i}^{*}.
 \end{cases}
\end{align*}

\hspace{0.2 in}Finally, the bias-corrected MSE estimators are defined as:
\begin{align}\label{mseadd}
\widehat{MSE}_{i}^{Add} & = \hat{M}_{1i}^{Add}(\hat{\bm{\psi}})+ \hat{M}_{2i,L} ,
\end{align}
\begin{align}\label{msemult}
\widehat{MSE}_{i}^{Mult} & =  \hat{M}_{1i}^{Mult}(\hat{\bm{\psi}})+ \hat{M}_{2i,L},
\end{align}
\begin{align}\label{msecomp}
    \widehat{MSE}_{i}^{Comp}&= 
    \begin{cases}
\widehat{MSE}_{i}^{Add} , & \text{if }\hat{M}_{1i}\geq \bar{M}_{1i}^{*}\\
\widehat{MSE}_{i}^{Mult}, & \text{if }\hat{M}_{1i} < \bar{M}_{1i}^{*},\\
    \end{cases}
\end{align}
and
\begin{align}\label{msehm} 
    \widehat{MSE}_{i}^{HM}&=
    \begin{cases}
    2\hat{M}_{1i} - \bar{M}_{1i}^{*}  + \hat{M}_{2i,L}, & \text{if }\hat{M}_{1i} \geq \bar{M}_{1i}^{*}\\
\hat{M}_{1i} exp\big[-\big\{\bar{M}_{1i}^{*}-\hat{M}_{1i}\big\}/\bar{M}_{1i}^{*}\big] + \hat{M}_{2i,L}, & \text{if }\hat{M}_{i}  < \bar{M}_{1i}^{*}.\\
    \end{cases}
\end{align}

We conclude the definition of the MSE estimator with a comment on computational efficiency. The bootstrap procedure of \cite{hall2006nonparametric} would require $ (BL)^{2} $ simulated samples. The fast double-bootstrap procedure of \cite{erciulescu2016small} would require $2BL$ simulated samples. In contrast, our proposed procedure only requires $BL$ simulated samples. Therefore, our procedure is more computationally efficient than competing methods in the literature.

\subsection{Confidence Interval }\label{sec2.5CI}

A benefit of the simulation-based procedure for constructing the predictor is that it naturally furnishes samples $\{\hat{\theta}_{i}^{(\ell)}: \ell = 1,\ldots, L\}$  from the estimated conditional density of the population parameter given the observed data.  One can use these samples to construct a prediction interval.   Observe that if $\bm{\psi}$ were known, then one could define a prediction interval with coverage rate of exactly $1-\alpha$ by $CI_{i}(\bm{\psi}) = (q_{\alpha/2}(\bm{y}_{si}, \bm{\psi}), q_{1-\alpha/2}(\bm{y}_{si}, \bm{\psi}))$, where $q_{\alpha}(\bm{y}_{si}, \bm{\psi})$ is the $\alpha$th quantile of $f_{p}(\theta_{i}\mid D_{s}, I_{i}; \bm{\psi})$ for the informative design and $q_{\alpha}(\bm{y}_{si}, \bm{\psi})$ is the $\alpha$th quantile of $f_{p}(\theta_{i}\mid \bm{y}_{si}, \bm{x}_{N_{i}}; \bm{\psi})$ for the noninformative design. In practice, $\bm{\psi}$ is unknown and must be estimated. Define an estimated prediction interval by 
\begin{align}\label{cinaive}
\widehat{CI}_{i}^{naive} = (q_{\alpha/2}(\bm{y}_{si}, \hat{\bm{\psi}}), q_{1-\alpha/2}(\bm{y}_{si}, \hat{\bm{\psi}})),
\end{align}
where $q_{\alpha}(\bm{y}_{si}, \hat{\bm{\psi}})$ is the $\alpha$th quantile of $\{\hat{\theta}_{i}^{(\ell)}: \ell = 1,\ldots, L\}$. It is widely known that the prediction interval (\ref{cinaive}) is has coverage less than $1-\alpha$ because it does not reflect variation associated with estimating $\bm{\psi}$ \citep{carlin1991sample}. 

We calibrate the interval (\ref{cinaive}) to obtain a prediction interval with improved coverage for estimated $\bm{\psi}$. To do this, we adapt the method of \cite{carlin1991sample} to the small area context. For any $\tilde{\bm{\psi}}$, let $q_{\alpha}(\bm{y}_{si}, \tilde{\bm{\psi}})$ be the $\alpha$th quantile of $\{\hat{\theta}_{i}^{(\ell)}(\tilde{\bm{\psi}}): \ell = 1,\ldots, L\}$, where  $\hat{\theta}_{i}^{(\ell)}(\tilde{\bm{\psi}})\sim f_{p}(\theta_{i} \mid D_{s}, I_{i}; \tilde{\bm{\psi}})$ for the informative design and  $\hat{\theta}_{i}^{(\ell)}(\tilde{\bm{\psi}})\sim f_{p}(\theta_{i} \mid \bm{y}_{si}, \bm{x}_{N_{i}};  \tilde{\bm{\psi}})$ .  Find $\alpha_{i}'$ such that
\begin{align*}
\frac{1}{B}\frac{1}{L}\sum_{b = 1}^{B}\sum_{\ell = 1}^{L} I[ q_{ \alpha_{i}'/2}(\bm{y}_{si}, \hat{\bm{\psi}}^{(b)}) \leq \hat{\theta}_{i}^{(\ell)} \leq q_{1-\alpha_{i}'/2}(\bm{y}_{si}, \hat{\bm{\psi}}^{(b)})] = 1-\alpha,
\end{align*}
where $\{\hat{\theta}_{i}^{( b)}\sim f_{p}(\theta_{i} \mid D_{s}, I_{i}, \hat{\bm{\psi}}): b = 1,\ldots, B\}$ for the informative design, $\{\hat{\theta}_{i}^{( b)}\sim f_{p}(\theta_{i} \mid \bm{y}_{si}, \bm{x}_{N_{i}}, \hat{\bm{\psi}}): b = 1,\ldots, B\}$ for the noninformative design, $\hat{\bm{\psi}}$ is the original estimator of $\bm{\psi}$, and $\{\hat{\bm{\psi}}^{(b)}: b = 1,\ldots, B\}$ are corresponding bootstrap estimates. Define the confidence interval by
\begin{align}\label{cibc}
\widehat{CI}_{i}^{Cal} &=[q_{ \alpha_{i}'/2}(\bm{y}_{si}, \hat{\bm{\psi}} ), q_{1- \alpha_{i}'/2}(\bm{y}_{si}, \hat{\bm{\psi}} )].   
\end{align}


\section{Simulations}\label{sec3Simulations}

We evaluate the proposed MSE estimator and confidence interval in the context of noninformative and informative sampling in Sections 3.1 and 3.2, respectively. 
The simulation model employed in the following studies is based on \cite{pfeffermann2007small} with a slight tweak in terms of sample design. That is, for each simulation $m=1,\hdots, M (=10000)$, we generate a finite population from the super-population model given by 
\begin{align}
y_{ij} = \beta_{0}+\beta_{1}{x}_{ij}+u_{i}+e_{ij}, \ j=1,\ldots,N_i, \ i=1,\ldots,D,
\end{align}
where $u_{i}\stackrel{iid}{\sim} N(0,\sigma^{2}_{u})$ and $e_{ij}\stackrel{iid}{\sim} N(0, \sigma^{2}_{e})$, truncated at $\pm2.5\sigma_{u}$ and $\pm2.5\sigma_{e}$, respectively. The coefficients are given by $(\beta_{0}, \beta_{1}) = (5, 0.1)$ and the covariates $x_{ij} \stackrel{iid}{\sim} U(0,1)$ are fixed over the whole simulations. Then, we stratify the areas into three strata $U_{h}$, $h=1,2,3$, where the sample size $n_{i}=5,10,$ and $15$, if $i\in U_{1},$ $U_{2}$, and $U_{3}$, respectively. A new sample is drawn in each MC iteration in accordance with the described sample design of each section.

We consider 6 types of area parameters. As a linear parameter, we consider the mean for area $i$, $\bar{Y}_{i}= N_{i}^{-1}\sum_{j=1}^{N_i}y_{ij}$. We also consider 5 non-linear parameters. The first one is the area mean of the exponentials of the $y_{ij}$ values, defined as $\exp_i = N_{i}^{-1}\sum_{i=1}^{N_i}\exp(y_{ij})$. The parameter $\exp_i$ is important because it represents the area mean in the common situation in which the data are modeled in the log scale. We then consider the  $25$th and $75$th quantiles. As in \cite{hyndman1996sample}, we define the $p$th quantile by $Q_{i, p}(y_{i1},\hdots,y_{iN_i}) = (1 - \omega)y_{i[j]} + \omega y_{i[j+1]},$ where $j = {\rm floor}(N_ip+1-p)$, $\omega = N_ip+1-p-j$, and $y_{i[j]}$ is the $j$th order statistic of $\{y_{i1},\hdots,y_{iNi}\}.$ This can be calculated through the function \texttt{quantile} in \texttt{R}. We then consider the poverty gap (PG) indicator defined as  
$${\rm PG}_i = \frac{1}{N_{i}}\sum_{j=1}^{N_i}\Big(\frac{z-\exp(y_{ij})}{z}\Big)I_{\big\{\exp(y_{ij})<z \big\}},$$ where $z = 155$. This value of $z$ was chosen to be roughly 0.6 times the median of $\exp(y_{ij})$ for a population. Finally, we define the Gini coefficient by $${\rm Gini}_i = \frac{\sum_{k=1}^{N_i}\sum_{\ell=1}^{N_i}\mid \exp(y_{ik}) -\exp( y_{i\ell})\mid}{2N_i^{2}\mbox{exp}_{i}}.$$ We calculate the Gini coefficient using the function \texttt{gini} of \texttt{R} package \texttt{reldist}.

\subsection{Noninformative Sampling}\label{sec3.1Noninfo}

In this simulation study, the four simulation configurations are considered by varying the number of areas as $D \in \{20,100\}$ and the ratio of variances as $R_{\sigma}=\sigma_{u}/\sigma_{e} \in \{1,2\}$, where $\sigma_{e}$ is fixed at $0.3$. We label the configurations $\{20,1\}$, $\{20,2\}$, $\{100,1\}$, and $\{100,2\}$ as 1, 2, 3, and 4, respectively. All areas are sampled and $n_{i}$ units within each area $i$ are sampled with simple random sampling. 

\begin{figure}[H]
    \centering
    \includegraphics[width=12cm]{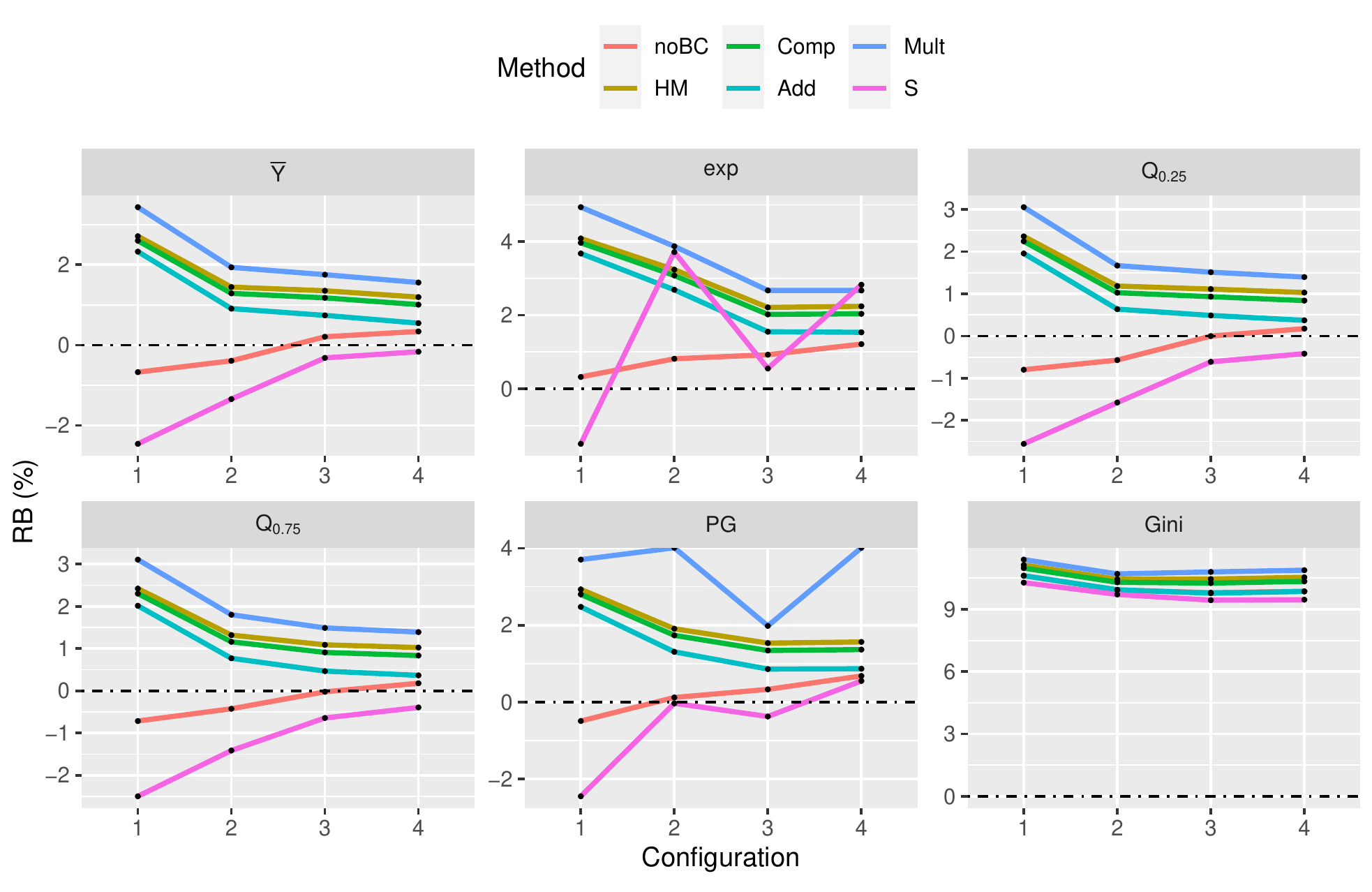}
    \caption{RBs of the proposed and standard (S) MSE estimators.}
    \label{noninfor_RB_MSE}
\end{figure}

Figure~\ref{noninfor_RB_MSE} displays the relative biases (RBs) of the  MSE estimators proposed in (\ref{MSEhatNoBC}), (\ref{mseadd}), (\ref{msecomp}),  (\ref{msemult}), and (\ref{msehm}), which are abbreviated noBC, Add, Comp, Mult, and HM, respectively. The standard MSE estimator (abbreviated ``S'') of \cite{molina2010small} is also presented. For a given configuration, the relative biases (RBs) in Figure \ref{noninfor_RB_MSE} for a MSE estimator $t\in \{\rm noBC, Add, Mult, Comp, HM, S\}$ is calculated by 
 \begin{align}\label{rbdef_noninfo} 
    {\rm RB}(\widehat{\rm MSE}_{t}) = 100 *\frac{(DM)^{-1}\sum_{i=1}^{D}\sum_{m=1}^{M}\widehat{\rm MSE}_{i,t}^{(m)} -  \rm MSE_{\rm DM}}{\rm MSE_{DM}},
    \end{align}
where ${\rm MSE_{DM}} = (DM)^{-1}\sum_{i=1}^{D}\sum_{m = 1}^{M}(\hat{\theta}_{i}^{(m)} - \theta_{i}^{(m)})^{2},$  
$\hat{\theta}_{i}^{(m)}$ is the predictor obtained in MC simulation $m$, $\theta_{i}^{(m)}$ is the true parameter generated in MC simulation $m$, and $\widehat{\rm MSE}_{i,t}^{(m)}$ is the type $t$ MSE estimator obtained in MC simulation $m$ for $t\in \{\rm noBC, Add, Mult, Comp, HM, S\}$.

In most cases, the absolute value of the RB decreases as the number of areas increases for a given $R_{\sigma}$. The MSE estimator without the bias correction (noBC) often has a absolute RB close to zero. When the noBC method has a negative bias, the bias corrections rectify this problem. The bias corrections can lead to conservative MSE estimates, with positive relative biases. Among the bias corrected estimators, Mult tends to produce the largest RB's and Add tends to produce the smallest RB's. The compromise MSE estimators have RB's that are between those of Add and Mult.  The single-bootstrap MSE estimator of \cite{molina2010small} (abbreviated  ``S'') is analogous to noBC in that neither takes the bias correction into account. With the exception of Exp and Gini, the standard procedure (S) tends to produce negative MSE estimates.  Except for Gini, noBC is preferred to S on the whole.

\begin{figure}[H]
    \centering
    \includegraphics[width=17cm]{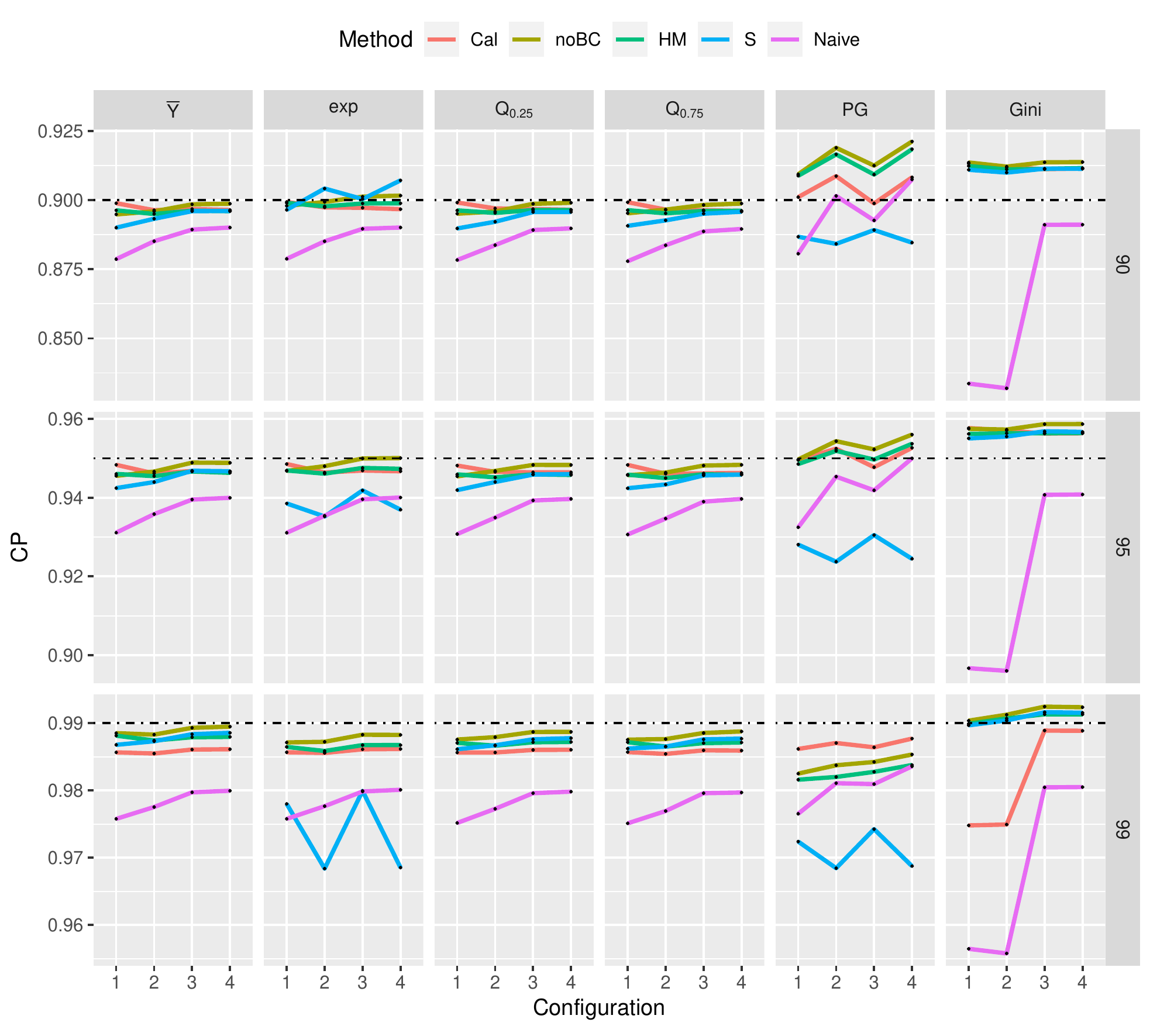}
    \caption{ECP of alternative prediction intervals for six parameters and three values of the nominal coverage probability.}
    \label{noninfor_CI_ECP}
\end{figure}

Figure~\ref{noninfor_CI_ECP} contains the empirical coverage probabilities (ECPs) of confidence intervals (CIs) defined in (\ref{cinaive}) and (\ref{cibc}) of Section \ref{sec2.5CI}, which are abbreviated ``Naive" and ``Cal", respectively. To calculate the empirical coverage probabilities (ECPs) for the normal-theory CIs, we consider the MSE estimators that assure the positive estimates. These MSE estimators are noBC, HM, and S. Because Comp shows the similar performance with HM, the ECPs of the normal-theory CI constructed with Comp are omitted in Figure \ref{noninfor_CI_ECP}. ECPs in Figure ~\ref{noninfor_CI_ECP} are obtained by $${\rm ECP}_{t}=(DM)^{-1}\sum_{i=1}^{D}\sum_{m=1}^{M}I\{\theta_{i}^{(m)}\in\widehat{\rm CI}_{i,t}^{(m)}\},$$ where $\widehat{\rm CI}_{i,t}^{(m)}$ ($t\in \{\rm no\_BC, HM, S, Naive, Cal\}$) is an interval estimate for $\theta_{i}$ at iteration $m$ . 

Both the normal-theory CI and calibrated CI almost obtain the nominal CP for $\bar{Y}$, $Q_{0.25}$, and $Q_{0.75}$. However, the CP of normal-theory CIs constructed with S is far less than the nominal CP, and is even less than that of Naive under certain configurations for exp and PG. For the PG, the calibrated CI works well. For Gini, the calibrated CI has a smaller CP under configuration 1 and 2, and it attains the nominal CP for configurations 3 and 4, where $D=100$.

\subsection{Informative Sampling}\label{sec3.2Info}

 We construct four scenarios by varying the ratio, $R_{\sigma}=\sigma_{e}^{-1}\sigma_{u}\in \{0.5, 1,2,3\}$.
 Stratum $U_1$, Stratum $U_2$, and Stratum $U_3$ contain areas $1\le i \le 50$, $51\le i\le 100$, and $101\le i\le 150$, respectively. Then, for each MC simulation, select $30$ areas from each stratum with probabilities $\pi_i=30 z_{i}/\sum_{j\in U_{h} }z_j$ by systematic sampling, where $z_{i}={\rm Round}[1000\times \exp(-u_i/8/\sigma_{u})].$ Sample $n_i$ units from the selected area $i$ with probabilities $\pi_{ij}=n_iz_{ij}/\sum_{k=1}^{N_i}z_{ik}$ by systematic sampling, where $z_{ij}=\exp\{[-(y_{ij}-\bx_{ij}\bm{\beta})/\sigma_{e}+\delta_{ij}/5]/3\}$, and $\delta_{ij}\sim N(0,1).$


\subsubsection{Results for Sampled Areas }

\hspace{0.2 in} Table~\ref{samp_MSE} shows the relative biases of the MSE estimators defined in Section \ref{sec2.3Informative} with the bias corrections of Section \ref{sec2.4BC}, as well as the relative bias of a standard parametric bootstrap MSE estimator (abbreviated ``S''). For each method $t$ and $R_{\sigma}$, the values in the table are given by 
 \begin{align}\label{rbdef} 
    {\rm RB}(\widehat{\rm MSE}_{t}) = 100 *\frac{\sum_{i=1}^{D}\widehat{\rm MSE}_{i,t}\Big/D -  \rm MSE_{\rm DM}}{\rm MSE_{DM}},
    \end{align}
where $\widehat{\rm MSE}_{i,t} =\sum_{m=1}^{M}A_{im}\widehat{\rm MSE}_{i,t}^{(m)}/\sum_{m=1}^{M}A_{im}$, ${\rm MSE_{DM}} = \sum_{i=1}^{D}{\rm MSE}_{i}\Big/{D},$  
\begin{align*}
{\rm MSE}_{i} = \sum_{m = 1}^{M}A_{im}(\hat{\theta}_{i}^{(m)} - \theta_{i}^{(m)})^{2}\big/\sum_{m = 1}^{M}A_{im},
\end{align*}
$\hat{\theta}_{i}^{(m)}$ is the predictor obtained in MC simulation $m$, $\theta_{i}^{(m)}$ is the true parameter generated in MC simulation $m$, and $\widehat{\rm MSE}_{i,t}^{(m)}$ is the type $t$ MSE estimator obtained in MC simulation $m$ with for $t\in \{\rm noBC, Add, Mult, Comp, HM, S\}$. The MSE estimators are defined in (\ref{nobcInf})-(\ref{msehm}) of Sections \ref{sec2.3Informative}-\ref{sec2.4BC}. The absolute relative biases of the proposed MSE estimators and S are usually below 5\% (and are uniformly below 10\%). The relative performances of the MSE estimators depend on the parameter. The standard (S) MSE estimator has slightly larger absolute relative biases than the proposed procedures for $\bar{Y}_{i}$, $Q_{0.75}$, and $\rm Exp$. For the other three parameters, the S procedure produces absolute relative biases that are similar to or below the absolute relative biases of the other methods. For $\rm PG$, the multiplicative bias-corrected MSE estimator produces infinity for a large $R_{\sigma}.$ When $R_{\sigma}$ is large, all units in a certain area could be smaller or larger than the poverty line $z$. In that case, the denominator of the multiplicative factor in (\ref{msemult})  could be 0, eventually making the relative bias infinity. Thus, ${\rm MSE}_{i}^{\rm HM}$ or ${\rm MSE}_{i}^{\rm comp}$ could be a good option when the variation among areas is large.



\begin{table}[H]
\captionsetup{font=footnotesize}
\centering
\caption{\label{samp_MSE} The relative biases (\%) of MSE estimators. Add, Mult, Comp, and HM are the bias-corrected MSE estimators.}
\scalebox{0.63}{
\begin{tabular}{llrrrr|llrrrr}
  \toprule
  \toprule
  \multirow{2}{*}{Parameter} & \multirow{2}{*}{Method} &\multicolumn{4}{c}{Scenario} &  \multirow{2}{*}{Parameter} & \multirow{2}{*}{Method} &\multicolumn{4}{c}{Scenario} \\
   \cmidrule{3-6} \cmidrule{ 9-12} 
      & & 0.5 & 1 & 2 & 3   & &  & 0.5 & 1 & 2 & 3 \\
  \midrule
 $\bar{Y}$ & noBC & -3.2888 & -1.6693 & -0.9937 & -0.7605 & $Q_{0.75}$ & noBC & -4.2644 & -2.6178 & -2.0532 & -1.8496 \\ 
   & Add & -3.1283 & -1.6218 & -1.0680 & -0.8543 &  & Add & -4.1483 & -2.5943 & -2.1464 & -1.9590 \\ 
   & Mult & -2.3594 & -0.8318 & -0.2733 & -0.0556 &  & Mult & -3.3858 & -1.8097 & -1.3565 & -1.1654 \\ 
   & Comp & -2.7739 & -1.2494 & -0.6849 & -0.4679 &  & Comp & -3.7939 & -2.2228 & -1.7642 & -1.5741 \\ 
   & HM & -2.6270 & -1.0950 & -0.5262 & -0.3079 &  & HM & -3.6471 & -2.0690 & -1.6060 & -1.4149 \\ 
   & S & -3.5099 & -1.9684 & -1.2815 & -1.0633 &  & S & -8.4663 & -6.7415 & -6.1197 & -5.9254 \\ \hline
 $\exp$ & noBC & -4.8626 & -2.8527 & -1.9576 & -1.4882 & $PG$ & noBC & 1.2872 & 0.7887 & 0.2677 & 0.0462 \\ 
   & Add & -4.8290 & -2.8991 & -2.1051 & -1.5930 &  & Add & 1.5305 & 0.9576 & 0.2876 & 0.0120 \\ 
   & Mult & -4.0221 & -2.0448 & -1.2328 & -0.7103 &  & Mult & 2.4300 & 1.8406 &   Inf &   Inf \\ 
   & Comp & -4.4504 & -2.4928 & -1.6830 & -1.1711 &  & Comp & 1.9361 & 1.3618 & 0.6932 & 0.4218 \\ 
   & HM & -4.2942 & -2.3257 & -1.5095 & -0.9977 &  & HM & 2.1028 & 1.5277 & 0.8596 & 0.5899 \\ 
   & S & -5.2376 & -3.2407 & -0.9142 & 8.0828 &  & S & 1.3097 & 0.4738 & 0.4789 & 0.5741 \\ \hline
  $Q_{0.25}$ & noBC & -2.4976 & -1.0116 & -0.3689 & -0.1270 & $Gini$ & noBC & 7.9220 & 8.0471 & 7.8479 & 7.5298 \\ 
   & Add & -2.3550 & -0.9628 & -0.4221 & -0.2029 &  & Add & 7.6900 & 7.8008 & 7.6066 & 7.2867 \\ 
   & Mult & -1.5628 & -0.1485 & 0.3980 & 0.6217 &  & Mult & 8.4909 & 8.6048 & 8.4075 & 8.0871 \\ 
   & Comp & -1.9886 & -0.5793 & -0.0288 & 0.1942 &  & Comp & 8.0870 & 8.2006 & 8.0051 & 7.6847 \\ 
   & HM & -1.8370 & -0.4207 & 0.1338 & 0.3583 &  & HM & 8.2512 & 8.3659 & 8.1699 & 7.8493 \\ 
   & S & 0.4553 & 1.7371 & 2.3741 & 2.5913 &  & S & 5.5700 & 5.4506 & 5.1709 & 4.9254 \\ 

   \bottomrule\bottomrule
\end{tabular}}
\end{table}

\begin{figure}[H]
\captionsetup{font=footnotesize}
    \centering
    \includegraphics[ height =9cm]{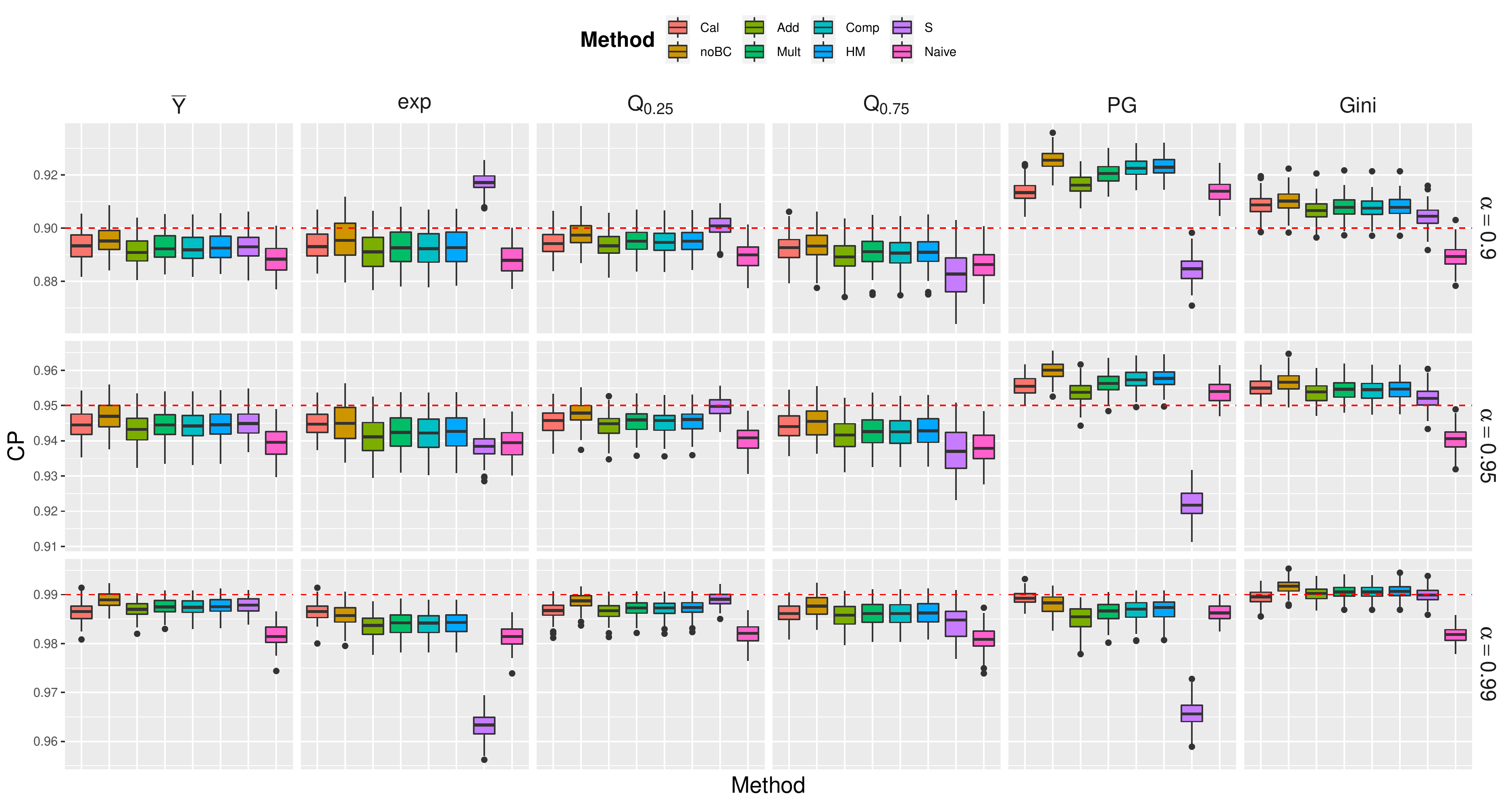}
    \caption{ \label{CP-boxplot} Boxplots of $\big\{ECP_{i}: i=1,\hdots,D \big\}$ for $R_{\sigma}=3$.}
 \end{figure}

\begin{figure}[H]
\captionsetup{font=footnotesize}
    \centering
    \includegraphics[height =10cm]{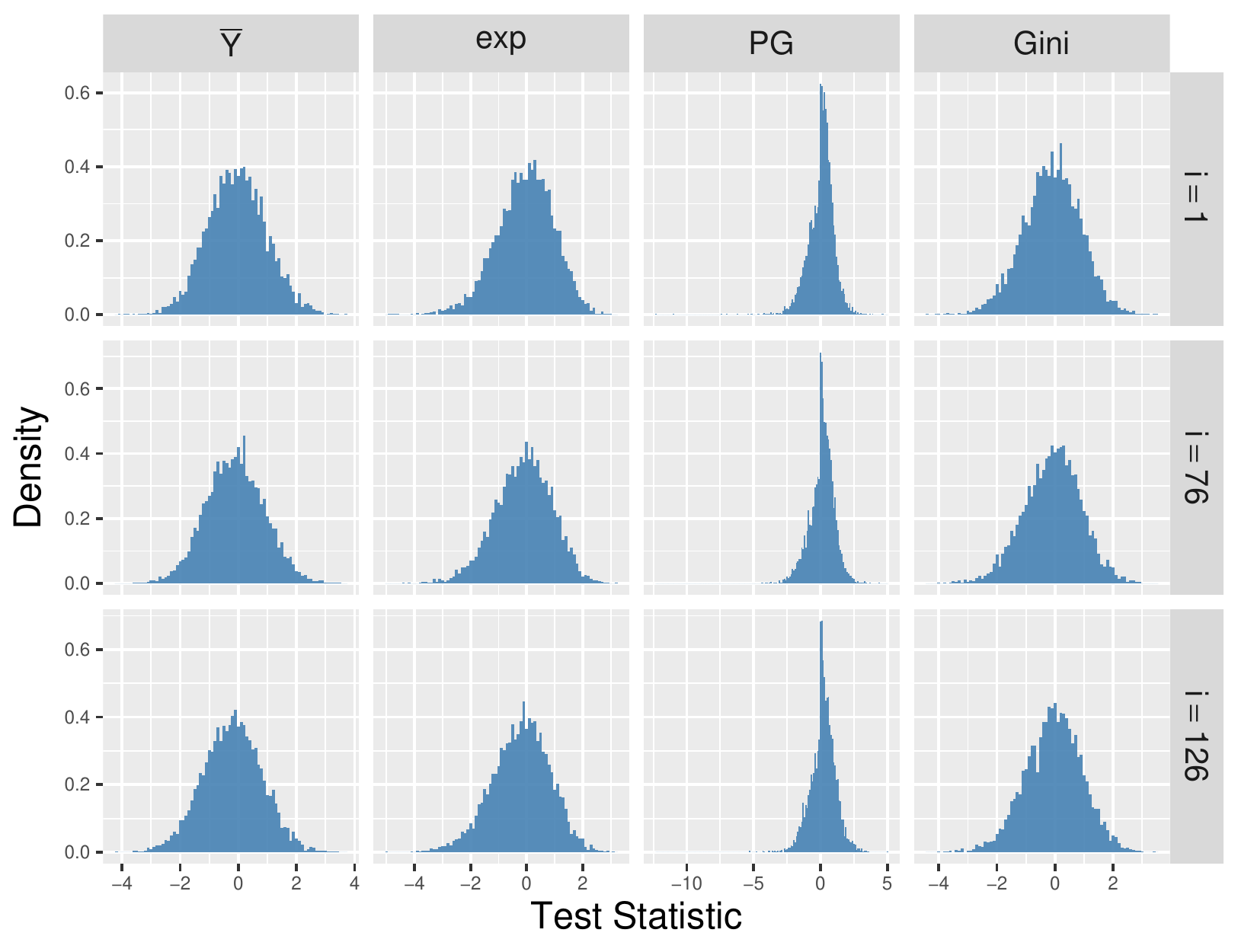}
    \caption{ \label{CP-fig} Histograms of $\big\{T_{i}^{(m)}, m=1,\hdots,10000 \big\}$ for $R_{\sigma}=3$, where the areas 50, 68, and 114 are randomly chosen for each stratum. The $HM$ bias correction of (\ref{msehm}) is used for the calculation of $T_{i}^{(m)}$.}
 \end{figure}

\hspace{0.2 in}Figure~\ref{CP-boxplot} shows boxplots of the empirical coverage probabilities given by $\{{\rm ECP}_{i}=\sum_{m=1}^{M}A_{im}I\{\theta_{i}^{(m)}\in\widehat{\rm CI}_{i}^{(m)}\}\big/{\sum_{m=1}^{M}A_{im}}: i = 1,\ldots, D\}$, where $\widehat{\rm CI}_{i}^{(m)}$ is an interval estimate for $\theta_{i}$ at iteration $m$. In Figure~\ref{CP-boxplot}, ``Naive'' and ``Cal'' denote the naive and calibrated confidence intervals defined in (\ref{cinaive}) and (\ref{cibc}) of Section \ref{sec2.5CI}. The remaining confidence intervals are normal theory confidence intervals given by 
\begin{align*}
    \widehat{\rm CI}_{{\rm Norm},i,t}^{(m)} = \hat{\theta}_{i}^{(m)} \pm z_{1-\alpha/2}\sqrt{\widehat{\rm MSE}_{i,t}^{(m)}},
\end{align*}
where $\hat{\theta}_{i}^{(m)}$ is the proposed predictor and $\widehat{\rm MSE}_{i,t}^{(m)}$ is the type $t$ MSE estimator ($t\in \{\rm no\_BC, Add, Mult, Comp, HM, S\}$) and obtained in simulation $m$. We present the coverage probabilities for $R_{\sigma} = 3$ in Figure~\ref{CP-boxplot}. The complete set of coverage probabilities is provided in Appendix E of the SM. While the naive confidence interval has the smallest coverage probability, the normal-theory confidence intervals almost attain the nominal coverage probability.  As shown in Figure \ref{CP-fig}, normal-theory confidence intervals may be inappropriate for some nonlinear parameters since the statistic
\begin{align*}
    T_i^{(m)} = \frac{\hat{\theta}_i^{(m)} - \theta_i^{(m)}}{\sqrt{\widehat{\rm MSE}_{i,t}^{(m)}}}
\end{align*}
does not have an approximately normal distribution. As seen in Figure \ref{CP-boxplot}, employing the calibrated confidence interval defined in Section \ref{sec2.5CI} may be a good alternative in this case.

\subsubsection{Results for Nonsampled Areas }

\begin{figure}[H]
\captionsetup{font=footnotesize}
    \centering
    \includegraphics[height = 9cm]{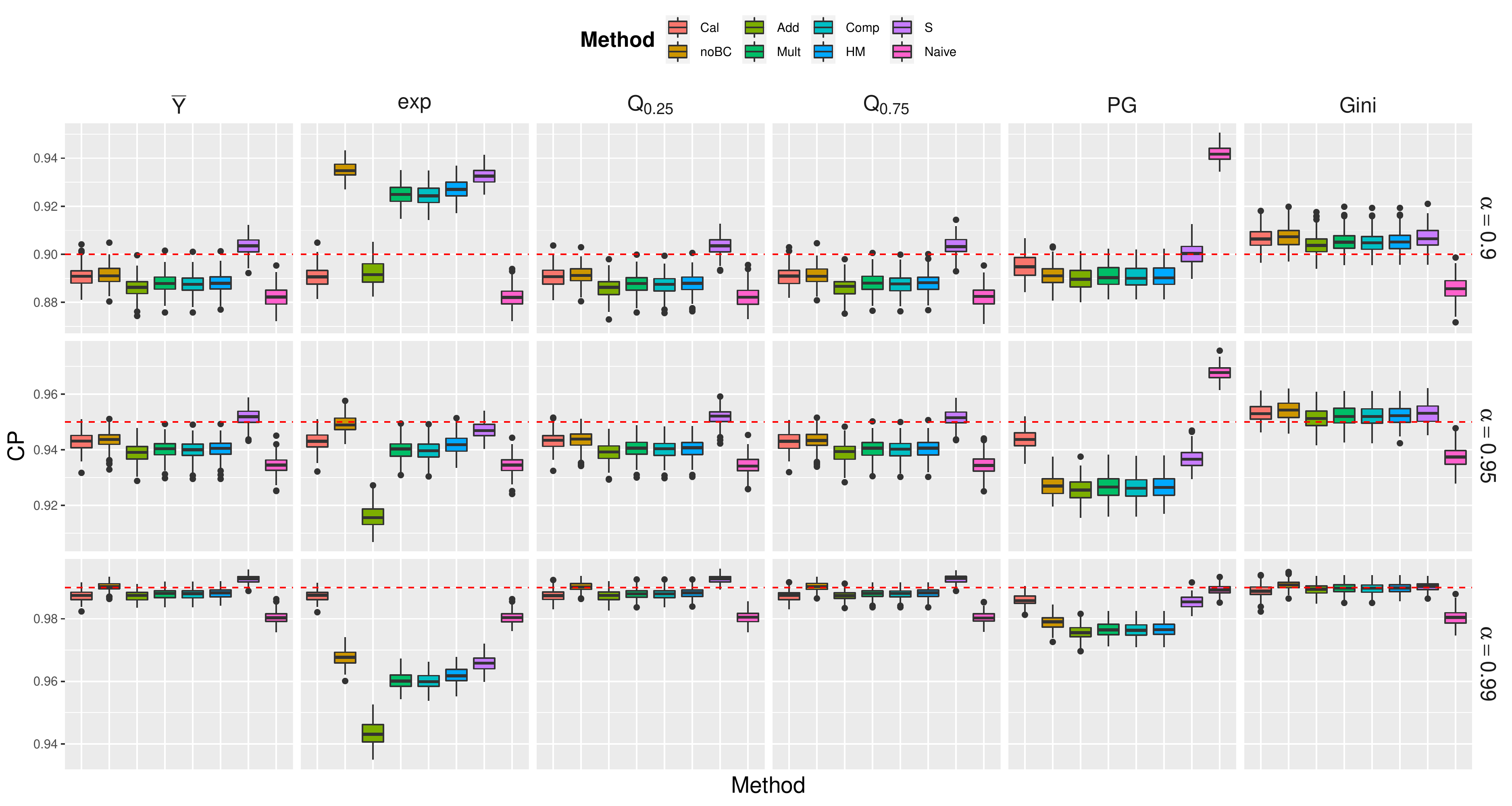}
    \caption{ \label{CP-boxplot-nonsamp} Boxplots of $\big\{{\rm ECP}_{i}: i=1,\hdots,D \big\}$ for $R_{\sigma}=3$ for non-sampled areas.}
 \end{figure}

\begin{figure}[H]
\captionsetup{font=footnotesize}
    \centering
    \includegraphics{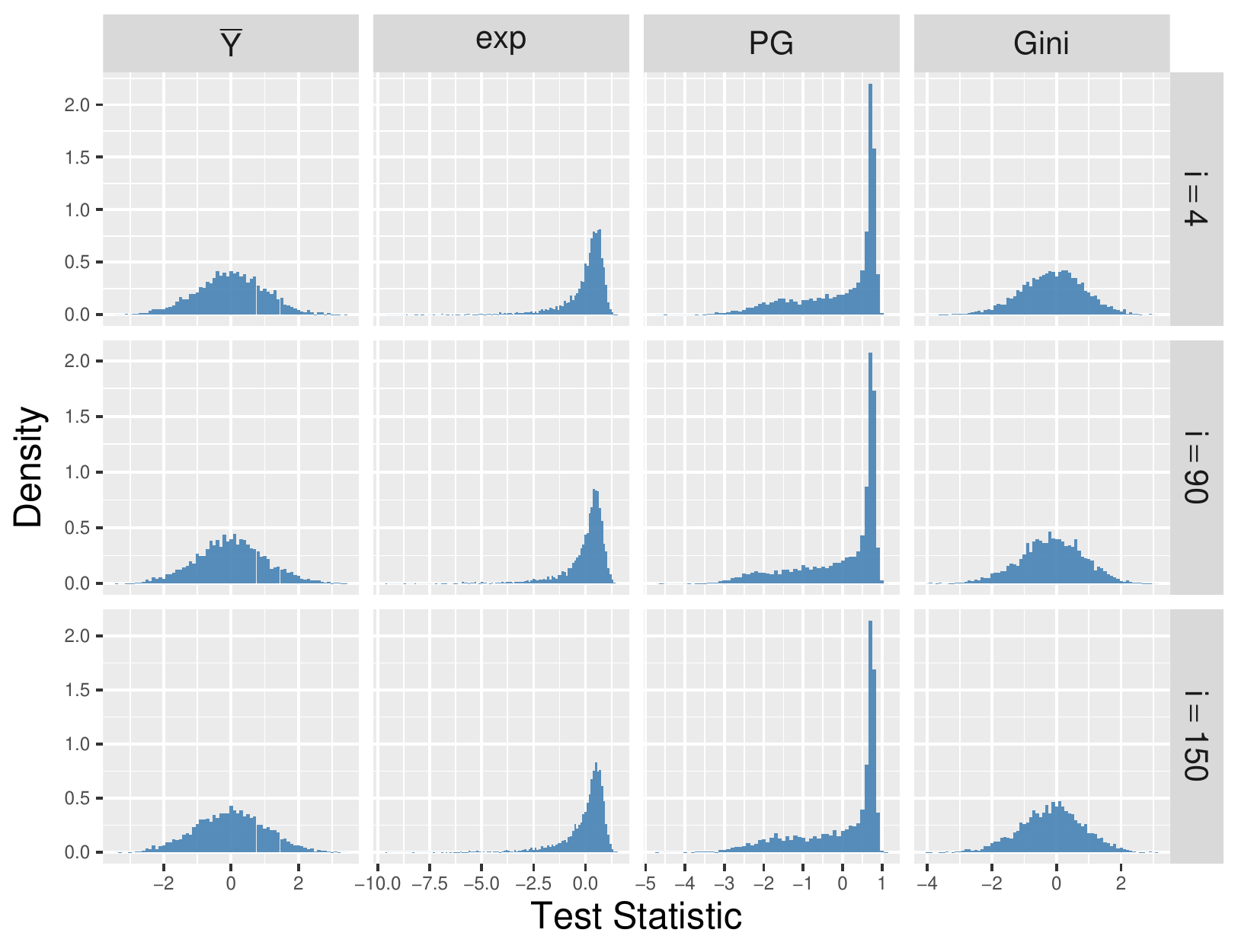}
    \caption{ \label{CP-fig-nonsamp} Histograms of $\big\{T_{i}^{(m)}: m=1,\hdots,10,000 \big\}$ for $R_{\sigma}=3$, where the areas 4, 90, and 150 are randomly chosen for each stratum. The $HM$ bias correction (\ref{msehm}) is used for the calculation of $T_{i}^{(m)}$.}
 \end{figure}

\hspace{0.2 in} Table \ref{nonsamp_MSE} and Figure \ref{CP-boxplot-nonsamp} give the average relative biases of the MSE estimators and the empirical coverages for nonsampled ares. The coverage probabilities in Figure~\ref{CP-boxplot-nonsamp} are for $R_{\sigma} = 3$, and the complete set of coverage probabilities is provided in in Appendix E of the SM.  The relative bias of an MSE estimator for a nonsampled area is defined as in (\ref{rbdef}), where ${\rm MSE}_{i} = \sum_{m = 1}^{M}(1-A_{im})(\hat{\theta}_{i}^{(m)} - \theta_{i}^{(m)})^{2}\big/\sum_{m = 1}^{M}(1-A_{im})$ and $\widehat{\rm MSE}_{i,t} =\sum_{m=1}^{M}(1 - A_{im})\widehat{\rm MSE}_{i,t}^{(m)}/\sum_{m=1}^{M}(1-A_{im})$ for $t\in \{\rm no\_BC, Add, Mult, Comp, HM, S\}$.  The empirical coverage probabilities are defined as $\{{\rm ECP}_{i}=\sum_{m=1}^{M}(1 - A_{im})I\{\theta_{i}^{(m)}\in\widehat{\rm CI}_{i}^{(m)}\}\big/{\sum_{m=1}^{M} (1 - A_{im}) }: i = 1,\ldots, D\}$, where $\widehat{\rm CI}_{i}^{(m)}$ is an interval estimate for $\theta_{i}$ at iteration $m$. The relative biases seem controlled well, except for ${\rm exp}$. It is beneficial to use the bias-corrected MSE estimators for ${\rm exp}$, $\rm PG$, and $\rm Gini$. Further, the proposed MSE estimators tend to have smaller absolute RBs compared to S. Although the bias-corrected MSE estimators are nearly unbiased, the normal theory confidence intervals can suffer from over-coverage or under-coverage. For Exp and PG, the normal theory CI's based on the bias-corrected MSE estimators can have coverage probabilities that are much larger (or smaller) than the nominal level. As illustrated in Figure~\ref{CP-fig-nonsamp}, the statistics $T_{i}^{(m)}$ can be very left-skewed for non-sampled areas. This can have adverse consequences for normal theory confidence intervals. We can see that the calibrated CI has stable empirical coverage probabilities for both exp and PG in Figure \ref{CP-boxplot-nonsamp}.


\begin{table}[!htbp]
\captionsetup{font=footnotesize}
\centering
\caption{\label{nonsamp_MSE} The relative biases (\%) of the proposed MSE estimators.}
\scalebox{0.75}{
\begin{tabular}{llrrrr|llrrrr}
  \toprule
  \toprule
  \multirow{2}{*}{Parameter} & \multirow{2}{*}{Method} &\multicolumn{4}{c}{Scenario} &  \multirow{2}{*}{Parameter} & \multirow{2}{*}{Method} &\multicolumn{4}{c}{Scenario} \\
     \cmidrule{3-6}\cmidrule{9-12}
      & & 0.5 & 1 & 2 & 3  & & &0.5 & 1 & 2 & 3 \\
  \midrule
$\bar{Y}$ & noBC & -1.3683 & -1.4256 & -1.1582 & -0.6466 & $Q_{0.75}$ & noBC & -1.5885 & -1.5482 & -1.1727 & -0.6586 \\ 
   & Add & -3.1419 & -2.4329 & -1.9914 & -1.4671 &  & Add & -3.3022 & -2.5476 & -2.0038 & -1.4777 \\ 
   & Mult & -2.3158 & -1.6337 & -1.1966 & -0.6673 &  & Mult & -2.4832 & -1.7502 & -1.2093 & -0.6783 \\ 
   & Comp & -2.6200 & -1.9805 & -1.5527 & -1.0275 &  & Comp & -2.7882 & -2.0966 & -1.5654 & -1.0384 \\ 
   & HM & -2.4065 & -1.7943 & -1.3720 & -0.8464 &  & HM & -2.5778 & -1.9110 & -1.3848 & -0.8574 \\ 
   & S & 6.8467 & 7.4575 & 7.9191 & 8.4831 &  & S & 6.1781 & 7.2045 & 7.8734 & 8.4575 \\ \hline
  $\exp$ & noBC & -0.2231 & 1.8354 & 10.8166 & 33.3885 & $PG$ & noBC & -1.9603 & -2.4381 & -2.2711 & -1.6499 \\ 
   & Add & -2.9583 & -0.7922 & 4.7869 & 17.1428 &  & Add & -2.7721 & -2.6968 & -2.2157 & -1.4249 \\ 
   & Mult & -1.9167 & 0.7558 & 10.2525 & 49.0299 &  & Mult & -1.6633 & -1.3688 & -1.0854 & -0.5383 \\ 
   & Comp & -2.2520 & 0.1543 & 7.7795 & 28.5723 &  & Comp & -2.1856 & -2.0457 & -1.6693 & -1.0012 \\ 
   & HM & -1.9693 & 0.5201 & 8.7725 & 31.4755 &  & HM & -1.9516 & -1.7919 & -1.4539 & -0.8302 \\ 
   & S & 6.8464 & 8.7585 & 16.2316 & 36.9442 &  & S & 8.1061 & 8.2381 & 7.8087 & 7.9402 \\ \hline
  $Q_{0.25}$ & noBC & -1.5697 & -1.4566 & -1.1708 & -0.6421 & $Gini$ & noBC & 6.1099 & 5.6732 & 6.0536 & 5.6795 \\ 
   & Add & -3.2915 & -2.4558 & -2.0039 & -1.4621 &  & Add & 5.8940 & 5.4628 & 5.8377 & 5.4697 \\ 
   & Mult & -2.4686 & -1.6572 & -1.2092 & -0.6622 &  & Mult & 6.6738 & 6.2427 & 6.6168 & 6.2475 \\ 
   & Comp & -2.7749 & -2.0043 & -1.5653 & -1.0225 &  & Comp & 6.2806 & 5.8480 & 6.2238 & 5.8548 \\ 
   & HM & -2.5636 & -1.8185 & -1.3846 & -0.8414 &  & HM & 6.4406 & 6.0074 & 6.3837 & 6.0140 \\ 
   & S & 6.5213 & 7.3836 & 7.8950 & 8.4803 &  & S & 6.9081 & 6.4332 & 6.8950 & 6.4982 \\ 
  \bottomrule\bottomrule
\end{tabular}}
\end{table}


\section{Data Analyses}

We illustrate the procedures through two data analyses. For the first, the design is noninformative for the model. For the second, we allow for the possibility of an informative sample design. 

\subsection{ Illustration with Noninformative Sample  }

We illustrate the proposed MSE estimator using the same data used in \cite{berg22database}. The data are from the 2019 Iowa Seat-Belt Use Survey. The areas are 15 sampled counties, and the units are the road segments in the sample for each county. Let $i = 1,\ldots, D$ index the counties and $j = 1,\ldots, n_{i}$ index the road segments in the sample for county $i$. Data collectors observe each sampled road segment for 45 minutes and record the proportion of vehicle occupants who are observed to be wearing a seat-belt. Let $p_{ij}$ denote the observed proportion for road segment $j$ in county $i$. 

As in Berg (2022), we specify a unit-level linear model for $y_{ij} = sin^{-1}(\sqrt{p}_{ij})$. The transformation is used to satisfy the normality assumption.  Berg (2022) uses a bivariate model for drivers and total occupants. For the purpose of this illustration, we simplify and only consider the univariate component corresponding to drivers. The model uses the two design variables that are used to select the road segments within counties. The first design variable is the road type. Let $R_{ij}$ denote the road type for road segment $j$ of county $i$, where $R_{ij}\in \{\mbox{Primary}, \mbox{Secondary}, \mbox{Local}\}$.  The second design variable, denoted by $v_{ij}$, is a measure of the annual average vehicle miles traveled across the road segment in a year. The model is given by 
\begin{align*}
y_{ij} = \beta_{0} + \bm{x}_{ij}'\bm{\beta}_{1} + b_{i} + e_{ij},
\end{align*}
where  $\bm{x}_{ij} = (I[R_{ij} = \mbox{Secondary}], I[R_{ij} = \mbox{Primary}])'$,  $b_{i}\stackrel{iid}{\sim}N(0,\sigma^{2}_{b})$ and $e_{ij}\stackrel{iid}{\sim}N(0, \sigma^{2}_{e}/v_{ij})$. Berg (2022) provides support for the model form as well as estimates of the model parameters. The design variable of road type is included in the model as a covariate, and Berg (2022) found that VMT was not significant as a covariate in the model. Because all significant design variables are included as model covariates, it is reasonable to assume that the design is non-informative for the model. 

The parameters of interest are the seat-belt use rates for the $D = 15$ counties in the sample.  The small area parameter of interest is defined as 
\begin{align*}
\theta_{i} = \frac{ \sum_{j = 1}^{N_{i}}  v_{ij}sin(p_{ij})^{2}  }{ \sum_{j = 1}^{N_{i}} v_{ij}  }.
\end{align*}
As a result of the transformation, the parameter is a nonlinear function of the model response variable. 

Berg (2022) presents estimates for the 15 sampled counties. As our focus is on mean square error estimation, we present the proposed mean square error estimates. Table~\ref{tabseatbelt} gives the estimates of the components of the MSE for the seat-belt survey application. The first column of Table~\ref{tabseatbelt} gives the estimates of the leading term. The second column gives the estimates of $M_{2i}$. The final column gives the percent of MSE  that is accounted for by the estimate of the bias of the estimate of the leading term. This is defined as $\widehat{Bias}_{i}/(\hat{M}_{1i} + \hat{M}_{2i,L})$, where we use the  additive estimate of the bias defined as $\widehat{Bias}_{i} = B^{-1}\sum_{b = 1}^{B}\hat{M}_{1i}^{(b)} - \hat{M}_{1i}$. The estimates of $M_{2i}$ are similar in magnitude to the estimates of $M_{1i}$ and can exceed the estimates of $M_{1i}$. For this data set, the sample size is small, and the variance due to parameter estimation is important.  Despite the small sample size, the contribution from the estimate of the bias of the estimator of the leading term in the MSE is negligible. The bias correction would lead to an increase in the estimated MSE for 8 out of the 15 counties and would lead to a decrease in the estimated MSE for the remaining 7 counties.

\begin{table}[H]
\centering
\caption{Estimates of components of MSE for seat-belt survey application.}\label{tabseatbelt}
\begin{tabular}{rrrr}
  \hline
 & $\hat{M}_{1i}$ & $\hat{M}_{2i}$ & $\frac{\widehat{Bias}_{i}}{(\hat{M}_{1i} + \hat{M}_{2i})}$ \\ 
  \hline
 1 & 0.01408 & 0.02403 & 0.00467 \\ 
 2 & 0.00136 & 0.02221 & 0.00274 \\ 
 3 & 0.00265 & 0.01770 & 0.00302 \\ 
 4 & 0.00953 & 0.02345 & -0.01396 \\ 
 5 & 0.00550 & 0.01711 & 0.01079 \\ 
 6 & 0.00416 & 0.01596 & 0.01530 \\ 
 7 & 0.01065 & 0.02179 & 0.00233 \\ 
 8 & 0.00110 & 0.01584 & -0.00060 \\ 
 9 & 0.00097 & 0.01972 & -0.00278 \\ 
 10 & 0.00480 & 0.01964 & 0.00207 \\ 
 11 & 0.00040 & 0.01963 & -0.00083 \\ 
 12 & 0.00124 & 0.01579 & -0.00241 \\ 
 13 & 0.00108 & 0.01851 & -0.00402 \\ 
 14 & 0.00205 & 0.02027 & -0.00024 \\ 
 15 & 0.00716 & 0.01462 & -0.02217 \\ 
   \hline
\end{tabular}
\end{table}

\subsection{Illustration with Informative Sample Design }

A nationwide survey of cropland is conducted as part of the Conservation Effects Assessment Project (CEAP) to examine the environmental impacts of conservation efforts on cropland. The sample for the CEAP survey is a subset of a larger survey called the National Resources Inventory (NRI). Data collection in the NRI uses aerial photographs of sampled units. More detailed information is collected for the CEAP subsample through farmer interviews. 

An important variable of interest in the CEAP and NRI surveys is sheet and rill erosion, which is soil loss due to the flow of water. The NRI measures sheet and rill erosion using a conventional model called the Universal Soil Loss Equation (USLE). CEAP obtains sheet and rill erosion using a more advanced computer model called the Revised Universal Soil Loss Equation (RUSLE2). USLE is measured for all units in the NRI survey. RUSLE2 is only available for the CEAP subsample because the RUSLE2 model requires detailed information that can only be gathered through farmer interviews. Our goal is to construct estimates of functions of RUSLE2, using the USLE as a covariate, for Iowa counties.

\begin{figure}[H]
\captionsetup{font=footnotesize}
    \centering
    \includegraphics[height = 8cm]{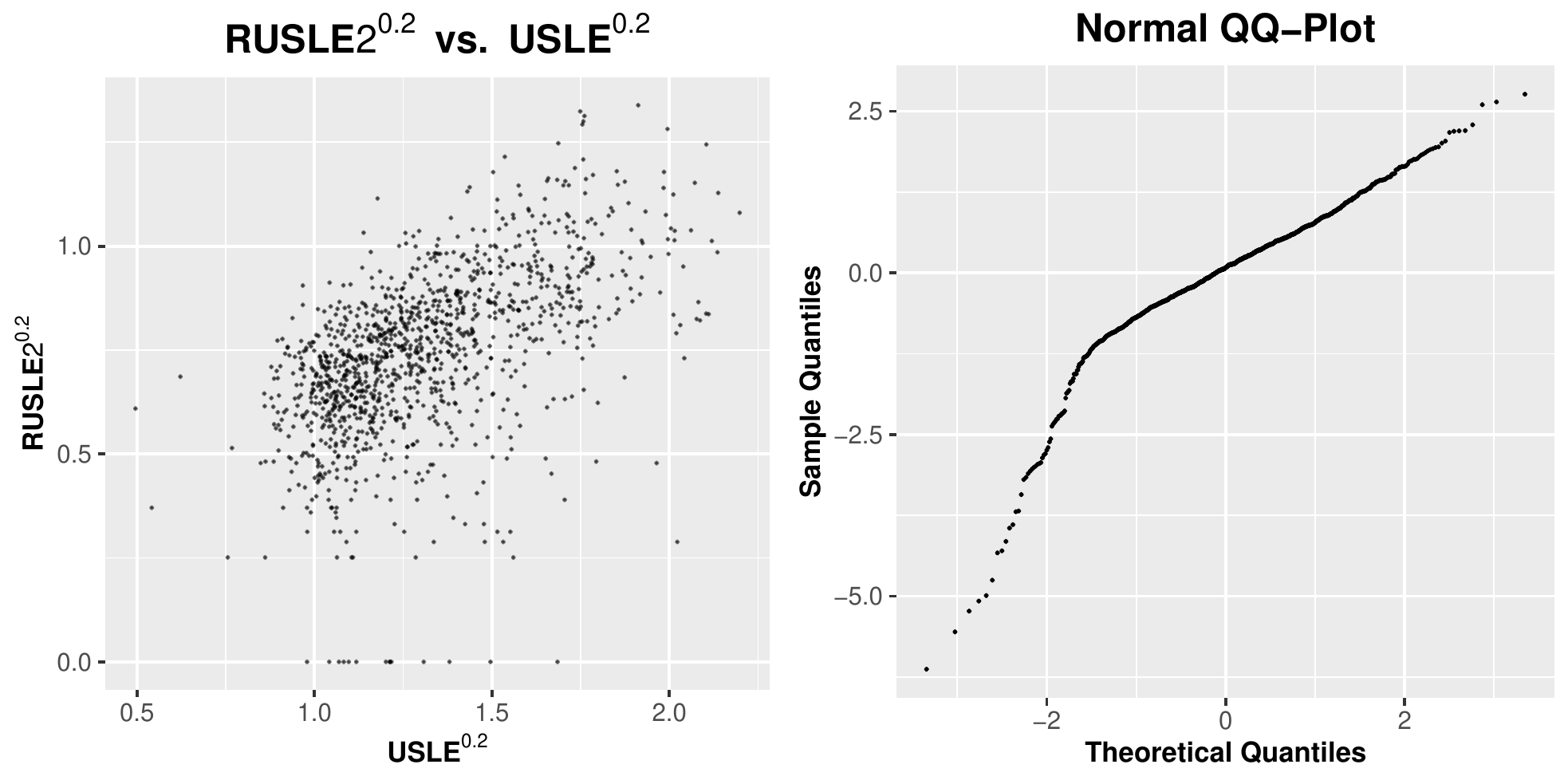}
    \caption{ \label{model_check} Scatter plot of the $0.2$ power transformed $RUSLE2$ and $USLE$ (Left), QQ normal plot for the generalized residuals (Right).}
 \end{figure}

Let $RUSLE2_{ij}$ and $USLE_{ij}$, respectively, denote the RUSLE and USLE for unit $j$ in county $i$. Motivated by the data analysis of \cite{berg2016imputation}, we transform these variables by a power of 0.2. This transformation not only yields the linear relationship seen in the left panel of Figure \ref{model_check} but also avoids problems with zeros that we would encounter with a log transformation. We then fit model unit-level linear model of \cite{battese1988error} to the sample data, where $y_{ij}$ and $x_{ij}$ are $RUSLE2_{ij}^{0.2}$ and $USLE_{ij}^{0.2}$, respectively, for $j$ point of county $i$, where $i=1,\hdots,99$, and $D = 99$ because Iowa has 99 counties. This model is used in \cite{molina2022nonlinearinformative} in the context of an informative sample design and is defined precisely in Appendix F of the SM. To assess the fitted model, we consider the generalized residuals ${r}_{ij}$, defined as
\begin{align*}
    {r}_{ij} =\int_{-\infty}^{y_{ij}}\frac{1}{\hat{\sigma}_{e}}\phi\bigg(\frac{y_{ij}-\hat{\beta}_{0}-\hat{\beta}_{1}x_{ij}-\hat{u}_{i}(\hat{\bm{\beta}},\hat{\sigma}_{u}^{2},\hat{\sigma}_{e}^{2})}{\hat{\sigma}_{e}}\bigg)dy_{ij}, \text{\hspace{0.2cm} $j=1,\hdots,n_{i}, i=1,\hdots,99$}.
\end{align*}
Under the assumed model, $\big\{{r}_{ij}:j=1,\hdots,n_{i}, i=1,\hdots,99\big\}$ behaves as like a sample from the uniform distribution. Thus, we compare the distribution of $\big\{\Phi^{-1}({r}_{ij}):j=1,\hdots,n_{i}, i=1,\hdots,99\big\}$ with the standard normal distribution through the QQ-plot in Figure \ref{model_check} (right). Due to 0 values in $y_{ij}$, the distribution of $\Phi^{-1}(r_{ij})$ looks more left-skewed than a normal distribution. However, the zeros only account for $1\%$ of the sample data, so we keep the fitted model.

Next, we regress $log(1/\pi_{ij})$ on $y_{ij}$, $x_{ij}$, and $x_{ij}y_{ij}$ with areas as fixed effects, to check if the interaction is significant. Since the coefficient for $x_{ij}y_{ij}$ is not significant with $p$-value $0.077$, we do not include an interaction between $y_{ij}$ and $x_{ij}$ in the weight model. We estimate the model parameters using the described in Appendix F of the SM, and we construct jackknife standard errors.  Table \ref{data_parest} shows the estimates of model parameters and corresponding standard errors. The USLE is a significant predictor of RUSLE2. The confidence interval for $\gamma_{2}$ contains zero, indicating that the design is noninformative for the model. Nonetheless, we think it is prudent to employ the procedure for informative sampling to guard against possible bias. 

In this study, we consider four county-level parameters of interests defined by  $\overline{RUSLE2}_{i} = N_{i}^{-1} \sum_{j=1}^{N_{i}}y_{ij}^{5}$, $Q^{RULSE2}_{i,0.25} = Q_{i,0.25}(y_{i1}^{5},\hdots,y_{iN_{i}}^{5}),$ $Q^{RULSE2}_{i,0.75} = Q_{i,0.75}(y_{i1}^{5},\hdots,y_{iN_{i}}^{5}),$ and $P_{i,m}^{RUSLE2} =N_{i}^{-1} \sum_{j=1}^{N_{i}}I(y_{ij}^{5}<m)$, where $m =0.232$ is the state sample median estimated from the observed $RUSLE2$.  The power of 5 converts the transformed RUSLE2 values to the original scale.  

\begin{figure}[H]
\captionsetup{font=footnotesize}
    \centering
    \includegraphics[height = 8cm]{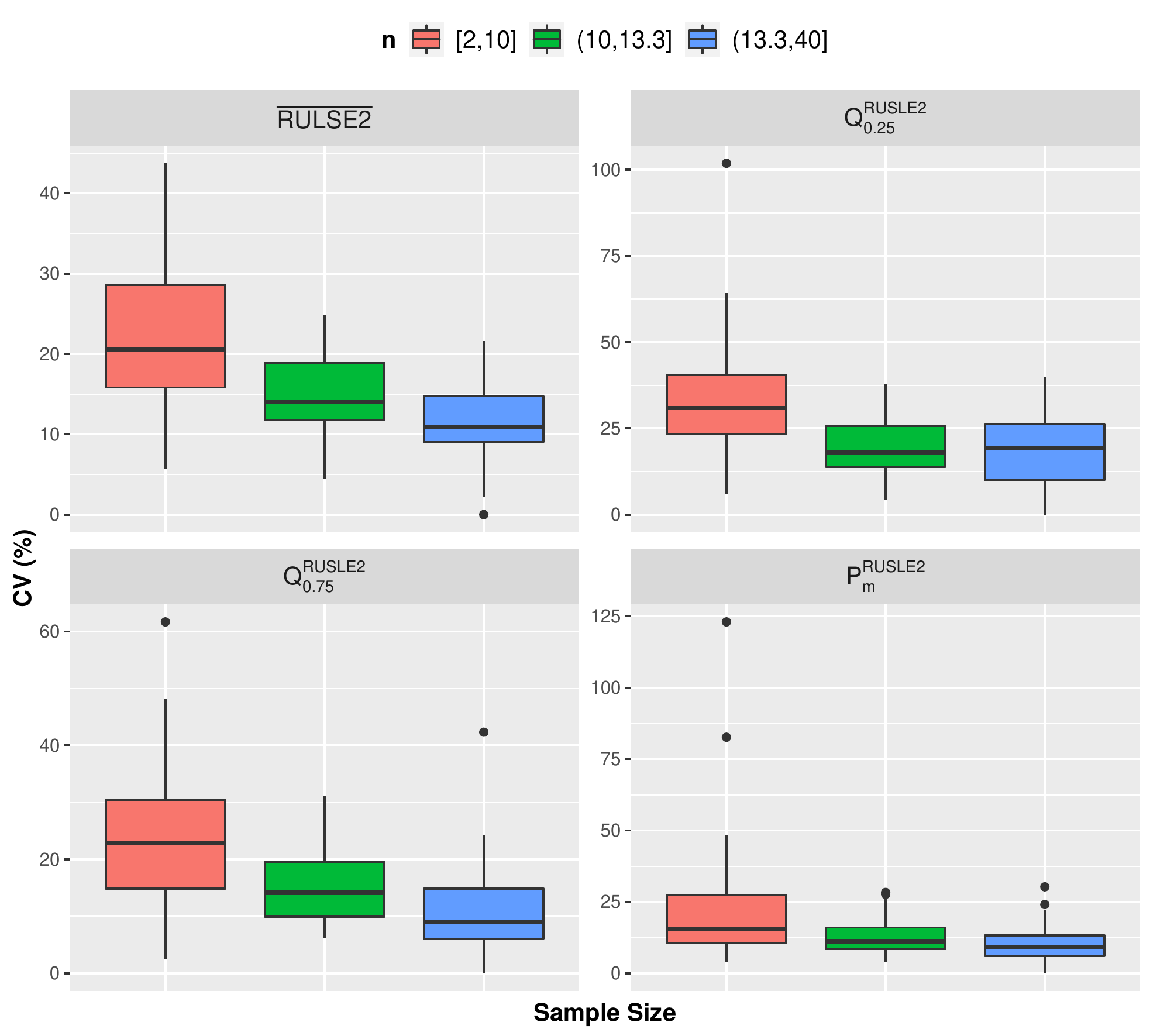}
    \caption{ \label{data_CV} Box plots of county's percent coefficients of variation (CV) grouped by sample sizes.}
 \end{figure}

\begin{figure}[H]
\captionsetup{font=footnotesize}
    \centering
    \includegraphics[width = 17.5cm]{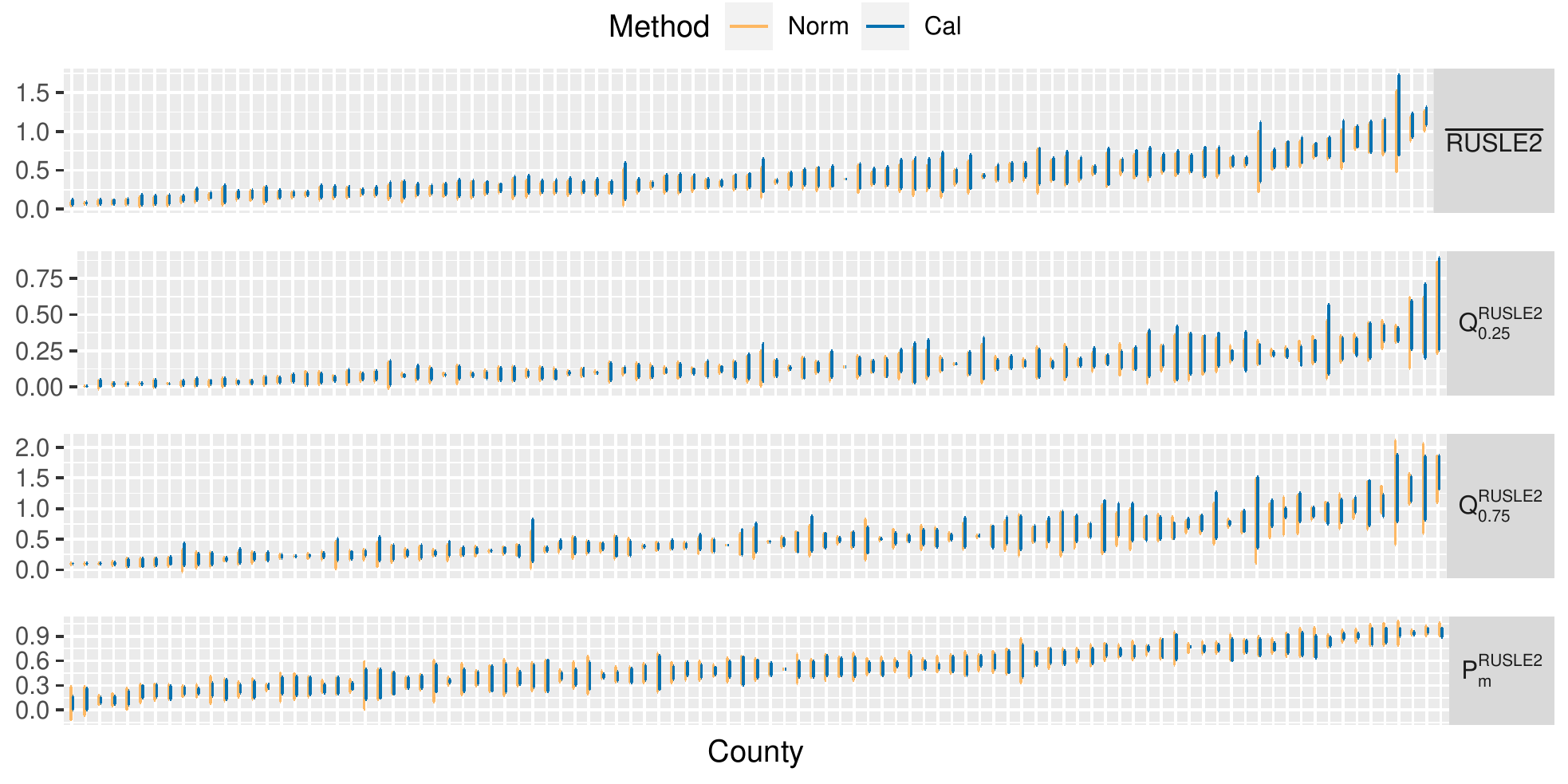}
    \caption{\label{data_CI}  Normal theory confidence intervals with $\widehat{MSE}_{i}^{HM}$ (Norm) and calibrated Confidence intervals (Cal), where counties are sorted by predicted values within each parameter.}
 \end{figure}

We construct predictors using the methods of \cite{molina2022nonlinearinformative}, and we use the procedures proposed in Section 2 to ascertain the uncertainty of the predictors. The percent coefficients of variation (CV) for a county $i$, calculated by ${\sqrt{\widehat{MSE}_{i}^{HM}}} \big /{\hat{\theta}_{i}}$, are given in Figure \ref{data_CV}. Note that the variation of the prediction may be affected by the sample size within a county, so CVs may have large values when the sample size is small. The mean of CVs across all counties for each parameter, $\overline{RUSLE2}_{i}$, $Q^{RULSE2}_{i,0.25}$, $Q^{RULSE2}_{i,0.75}$, and $P_{i,m}^{RUSLE2}$, is $16.31\%$, $24.56\%$ $17.38\%$, and $16.27\%$, respectively. For a given county parameter, the normal theory confidence intervals with $\widehat{MSE}_{i}^{HM}$ and the calibrated confidence intervals are shown in Figure \ref{data_CI}. For $\overline{RUSLE2}_{i}$, there is no significant difference between lengths from two confidence intervals, but for the other county parameters, the normal theory confidence intervals are longer than that the calibrated confidence intervals.  

 \begin{table}[H]
\centering
 \caption{\label{data_parest}The estimates of model parameters and its confidence interval using the standard error from the jackknife variance estimates.}
\begin{tabular}{ccccc}
  \hline \hline
Parameter  & Estimates&SE& LL & UL \\ \hline
$\beta_{0}$ & 0.2219 & 0.0276 & 0.1678 & 0.2760 \\ 
  $\beta_{1}$ & 0.4006 & 0.0215 & 0.3584 & 0.4428 \\ 
  $\sigma_{u}^{2}$ & 0.0064 & 0.0012 & 0.0040 & 0.0088 \\ 
  $\sigma_{e}^{2}$ & 0.0205 & 0.0022 & 0.0163 & 0.0247 \\ 
  $\gamma_{1}$ & -0.1236 & 0.0487 & -0.2191 & -0.0281 \\ 
  $\gamma_{2}$ & -0.1215 & 0.0626 & -0.2441 & 0.0012 \\  \hline\hline
\end{tabular}
\end{table}


\begin{figure}[H]
\captionsetup{font=footnotesize}
    \centering
    \includegraphics[width = 12cm]{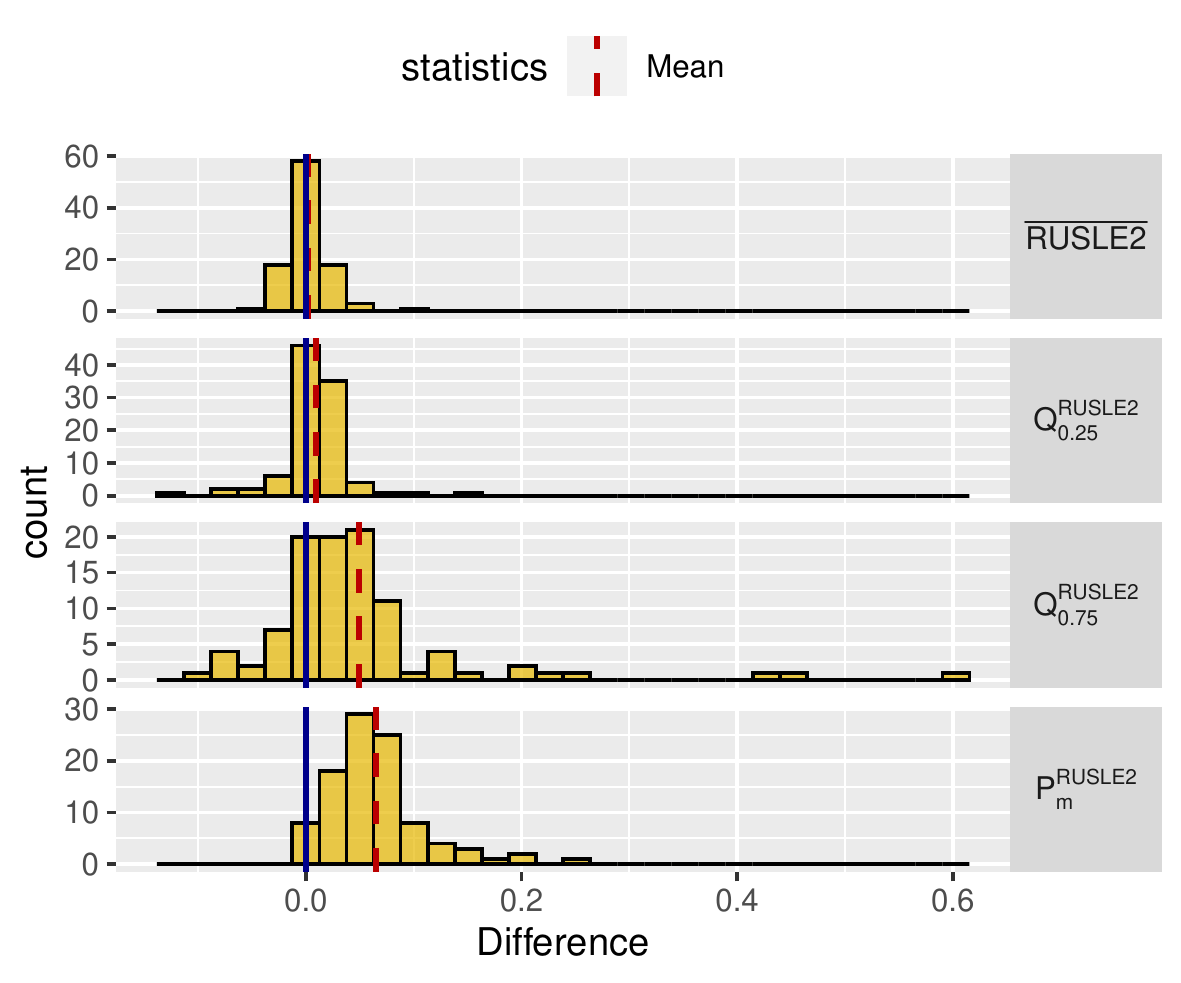}
    \caption{ \label{data_his_CI_diff} Histograms of differences between the lengths of mormal theory confidence intervals (with $\widehat{MSE}_{i}^{HM}$) and calibrated confidence intervals.}
 \end{figure}

\section{Conclusion}


We propose a procedure for conducting inference for general small area parameters for both a sampled and non-sampled area under an informative design. The proposed predictors are MC approximations  for optimal predictors under specified sample distribution models. For MSE estimation, we avoid using a fully parametric bootstrap by generating bootstrap parameter estimates from an estimate of the asymptotic normal distribution of the parameter estimators. We also construct estimates of the bias of the estimator of the leading term without use of the double bootstrap. Further, we define calibrated confidence intervals that do not require normal theory. 

The simulation study supports the proposed procedures. Both the pseudo-EB predictor and the proposed predictor are more efficient than predictors that ignore the design. The proposed procedure renders smaller mean square errors than the pseudo-EB predictor. We think that the efficiency gains from the proposed procedure, relative to the pseudo-EB predictor, occur because proposed procedure incorporates more information through stronger assumptions about the survey weight.  The proposed predictor is nearly as efficient for the mean as the optimal predictor of \cite{pfeffermann2007small}. 

We apply the proposed procedure to predict several functions of sheet and rill erosion (RUSLE2) for Iowa counties. In the data analysis, the calibrated prediction intervals are often narrower than normal theory prediction intervals, although both sets of intervals have similar coverage rates in the simulation study. A limitation of the data analysis is that we do not have covariates for a full population. Instead, we use covariates from a larger survey that is a super-set of the CEAP sample. Our procedure does not account for uncertainty due to lack of population-level auxiliary information. Accounting for this source of variation is a possible future research direction.

\bibliography{MSEGen1Refs}
\appendix

\section{ Discussion of SUMCA Method}

We consider the SUMCA MSE estimator of \cite{jiang2020sumca} for the EBP of \cite{molina2010small}. To emphasize that $\hat{M}_{1i}$ is a function of both the data and the parameter estimator, we express the estimator of the leading term as $\hat{M}_{1i} = M_{1i}(\bm{y}_{i}, \hat{\bm{\psi}})$. By equations (9), (11), and (13) of \cite{jiang2020sumca}, the SUMCA MSE estimator for the EBP is given by 
\begin{align}\label{sumcamse}
\hat{M}_{1i} + K^{-1}\sum_{k =1}^{K}\left\{     M_{1i}(\bm{y}_{i}^{(b)}, \hat{\bm{\psi}}) - M_{1i}(\bm{y}_{i}^{(b)}, \hat{\bm{\psi}}^{(b)})\right\}.
\end{align}
The SUMCA MSE estimator (\ref{sumcamse}) omits the estimator of $M_{2i}$. Our experience is that $M_{2i}$ is important for nonlinear parameters. Another potential problem with the SUMCA MSE estimator is that the additive form of the bias correction can produce negative MSE estimates, although \cite{jiang2020sumca} explain that the probability of a negative MSE estimate decreases as the sample size increases. As a result of these issues with the SUMCA MSE estimator for nonlinear parameters, we do not consider the SUMCA MSE estimator. 

\section{Proof of {\it Theorem 1} }

A justification for the MSE decomposition in (\ref{msedecomp1}) is as follows:
\begin{align*}
E[(\hat{\theta}_{i}^{EBP} - \theta_{i})^{2}\mid \bm{y}_{si}] & = E[(\hat{\theta}_{i}^{EBP} - \hat{\theta}_{i}^{EB} + \hat{\theta}_{i}^{EB} - \theta_{i})^{2}\mid \bm{y}_{si}]  \\ \nonumber
													  & = E[(\hat{\theta}_{i}^{EBP} - \hat{\theta}_{i}^{EB})^{2}\mid \bm{y}_{si}] + E[(\hat{\theta}_{i}^{EB} - \theta_{i})^{2}\mid \bm{y}_{si}] \\ \nonumber   & + 2E[(\hat{\theta}_{i}^{EBP} - \hat{\theta}_{i}^{EB})(\hat{\theta}_{i}^{EB} - \theta_{i})\mid \bm{y}_{si}]. 
\end{align*}
If we condition on $\theta_{i}$ and $\bm{y}_{si}$, then the only random variable remaining in the cross-term is $\hat{\theta}_{i}^{EBP}$, which depends on the $L$ simulated $\{\hat{\theta}_{i}^{(\ell)}: \ell = 1,\ldots, L\}$. We can therefore use a double expectation argument to show that the cross-term is zero, as 
\begin{align*}
 E\{E[(\hat{\theta}_{i}^{EBP} - \hat{\theta}_{i}^{EB})\mid \theta_{i}, \bm{y}_{si}](\hat{\theta}_{i}^{EB} - \theta_{i}) \} = 0.
\end{align*}

\section{ Proof of {\it Theorem 2} }

By construction, $\hat{\theta}_{i}^{(1)},\ldots, \hat{\theta}_{i}^{(L)}$ are $iid$, $ E_{L}[\hat{\theta}_{i}^{(\ell)}\mid  \bm{y}_{si}, \bm{x}_{N_{i}}; \hat{\bm{\psi}}] = E[\theta_{i} \mid \bm{y}_{si}, \bm{x}_{N_{i}}; \hat{\bm{\psi}} ],$
and $V_{L}(\hat{\theta}_{i}^{(r)}\mid \bm{y}_{si}, \bm{x}_{N_{i}}; \hat{\bm{\psi}}) = V(\theta_{i} \mid \bm{y}_{si}, \bm{x}_{N_{i}}; \hat{\bm{\psi}})$, where $V_{L}$ denotes variance relative to the distribution used to generate $\hat{\theta}_{i}^{(1)},\ldots, \hat{\theta}_{i}^{(L)}$. The estimator $\hat{M}_{1i}$ is the sample variance of $\hat{\theta}_{i}^{(1)},\ldots, \hat{\theta}_{i}^{(L)}$.  Result  (i) follows from standard properties of sample variances of $iid$ random variables. Further, by standard properties of the sample variance, $|\hat{M}_{1i} - V\{\theta_{i} \mid  \bm{y}_{si}, \bm{x}_{N_{i}}; \hat{\bm{\psi}} \}| = O_{p}(L^{-0.5})$. By Chebyshev's inequality, 
\begin{align*}
P(|V(\theta_{i} \mid \bm{y}_{si}, \bm{x}_{N_{i}}; \hat{\bm{\psi}}) -  V(\theta_{i} \mid \bm{y}_{si}, \bm{x}_{N_{i}}; \bm{\psi})| > \epsilon) \leq E[|V(\theta_{i} \mid \bm{y}_{si}, \bm{x}_{N_{i}}; \hat{\bm{\psi}}) -  V(\theta_{i} \mid \bm{y}_{si}, \bm{x}_{N_{i}}; \bm{\psi})|^{2}]/\epsilon^{2}.  
\end{align*}
By Taylor's theorem, 
\begin{align*}
E[|V(\theta_{i} \mid \bm{y}_{si}, \bm{x}_{N_{i}}; \hat{\bm{\psi}}) -  V(\theta_{i} \mid \bm{y}_{si}, \bm{x}_{N_{i}}; \bm{\psi})|^{2}] = E[(D_{1}(\bm{\psi}^{*}, \bm{y}_{si})(\hat{\bm{\psi}} - \bm{\psi}) )^{2}].  
\end{align*}
By assumption, $E[(D_{1}(\psi, \bm{y}_{si})(\hat{\bm{\psi}} - \bm{\psi}) + D_{2}(\bm{\psi}^{*}, \bm{y}_{si})(\hat{\bm{\psi}} - \bm{\psi})^{2})^{2}]  = o(1)$. Therefore, $|V\{\theta_{i} \mid \bm{y}_{si}, \bm{x}_{N_{i}}; \hat{\bm{\psi}}\} - V\{\theta_{i} \mid  \bm{y}_{si}, \bm{x}_{N_{i}}; \bm{\psi}\}|\stackrel{p}{\rightarrow} 0$ as $D \rightarrow \infty$.  By the triangle inequality, $|\hat{M}_{1i} -  V\{\theta_{i} \mid \bm{y}_{si}, \bm{x}_{N_{i}}; \bm{\psi}\}|\leq |\hat{M}_{1i} -  V\{\theta_{i} \mid \bm{y}_{si}, \bm{x}_{N_{i}}; \hat{\bm{\psi}} \}| + |V\{\theta_{i} \mid \bm{y}_{si}, \bm{x}_{N_{i}}; \hat{\bm{\psi}}\} - V\{\theta_{i} \mid \bm{y}_{si}, \bm{x}_{N_{i}}; \bm{\psi}\}|$, and result (ii) follows. 

\section{  Proof of {\it Theorem 3}}

The proof of Theorem 3 requires the following regularity conditions:
\begin{enumerate}
    \item [R1:] Let $$\ddot{\bm{G}}(\bm{\psi}, \bm{y}_{si}, \bm{x}_{N_{i}}) = \frac{ \partial^{2}}{\partial\bm{\psi} \partial \bm{\psi}'} g(\bm{\psi}, \bm{y}_{si}, \bm{x}_{N_{i}}) .$$ Assume $\ddot{\bm{G}}( \bm{\psi}, \bm{y}_{si}, \bm{x}_{N_{i}}) = O_{p}(1)$ for all $\bm{\psi}$ in the parameter space. 
    \item[R2:] Assume $\hat{\bm{\psi}}^{(b)} - \hat{\bm{\psi}} = O_{p}(D^{-0.5})$.
    \item[R3:] Assume 
    \begin{align*}
        \frac{D}{B}\sum_{b= 1}^{B} (\hat{\bm{\psi}}^{(b)} - \hat{\bm{\psi}})(\hat{\bm{\psi}}^{(b)} - \hat{\bm{\psi}})' = D Var(\hat{\bm{\psi}}) + o_{p}(1). 
    \end{align*}
\end{enumerate}
We now state the proof of Theorem 3.

{\it Proof of Theorem 3:} By definition, 
\begin{align*}
    \hat{M}_{2i,\infty} = \frac{1}{B}\sum_{b = 1}^{B} (g(\hat{\bm{\psi}}^{(b)} , \bm{y}_{si}, \bm{x}_{N_{i}}) - g(\hat{\bm{\psi}}, \bm{y}_{si}, \bm{x}_{N_{i}}))^{2}. 
\end{align*}
A second order Taylor expansion of $g(\hat{\bm{\psi}}^{(b)} , \bm{y}_{si}, \bm{x}_{N_{i}})$ around $g(\hat{\bm{\psi}}, \bm{y}_{si}, \bm{x}_{N_{i}})$ gives 
\begin{align*}
   \hat{M}_{2i,\infty}  = \frac{1}{B}\sum_{b = 1}^{B}( \dot{\bm{g}}(\hat{\bm{\psi}}, \bm{y}_{si}, \bm{x}_{N_{i}})'(\hat{\bm{\psi}}^{(b)} - \hat{\bm{\psi}}) + 0.5 (\hat{\bm{\psi}}^{(b)} -\hat{ \bm{\psi}})'\ddot{\bm{G}}(\tilde{\bm{\psi}}^{(b)}, \bm{y}_{si}, \bm{x}_{N_{i}})(\hat{\bm{\psi}}^{(b)} - \hat{\bm{\psi}}))^{2},
\end{align*}
where $\tilde{\bm{\psi}}^{(b)}$ is on the line segment joining $\hat{\bm{\psi}}^{(b)}$ and $\hat{\bm{\psi}}$. Conditions R1-R3 then imply that 
\begin{align*}
      \hat{M}_{2i,\infty} & =  \dot{\bm{g}}(\hat{\bm{\psi}}, \bm{y}_{si}, \bm{x}_{N_{i}})'\left[\frac{1}{B}\sum_{b =1}^{B}(\hat{\bm{\psi}}^{(b)} - \hat{\bm{\delta}})(\hat{\bm{\psi}}^{(b)} - \hat{\bm{\psi}})'   \right]\dot{\bm{g}}(\hat{\bm{\psi}}, \bm{y}_{si}, \bm{x}_{N_{i}}) + O_{p}\left( \frac{1}{D\sqrt{D}}\right) \\ \nonumber 
                & = \dot{\bm{g}}(\hat{\bm{\psi}}, \bm{y}_{si}, \bm{x}_{N_{i}})'Var(\hat{\bm{\psi}}) \dot{\bm{g}}(\hat{\bm{\psi}}, \bm{y}_{si}, \bm{x}_{N_{i}}) + o_{p}(D^{-1}).
\end{align*}
\qedsymbol 

\section{Complete Set of Coverage Probabilities for Informative Sample Design}

\begin{table}[H]
\captionsetup{font=footnotesize}
\centering
\caption{\label{samp_CI} The coverage probabilities of confidence intervals for sampled areas. Naive and Cal are the naive and the calibrated confidence interval, respectively. The remainder in the Method column is the normal theory confidence interval with the corresponding MSE estimator.}
\scalebox{0.52}{
\begin{tabular}{lllrrrrlrrrrlrrrr}
  \toprule
    \toprule
     \multirow{2}{*}{Parameter} & \multirow{2}{*}{Method} & &\multicolumn{4}{c}{$CP = 0.90$}& & \multicolumn{4}{c}{$CP = 0.95$}& & \multicolumn{4}{c}{$CP = 0.99$} \\
     \cmidrule{4-7} \cmidrule{9-12} \cmidrule{14-17}
    &  & &0.5 & 1 & 2 & 3 && 0.5 & 1 & 2 & 3&& 0.5 & 1 & 2 & 3  \\   \midrule
   $\bar{Y}$ & Cal &  & 0.890 & 0.893 & 0.894 & 0.893 &  & 0.942 & 0.944 & 0.945 & 0.945 &  & 0.985 & 0.986 & 0.986 & 0.986 \\ 
   & noBC &  & 0.891 & 0.895 & 0.896 & 0.895 &  & 0.944 & 0.946 & 0.947 & 0.947 &  & 0.988 & 0.989 & 0.989 & 0.989 \\ 
   & Add &  & 0.888 & 0.891 & 0.892 & 0.891 &  & 0.940 & 0.943 & 0.944 & 0.943 &  & 0.986 & 0.987 & 0.987 & 0.987 \\ 
   & Mult &  & 0.889 & 0.892 & 0.893 & 0.893 &  & 0.942 & 0.944 & 0.945 & 0.944 &  & 0.986 & 0.987 & 0.987 & 0.988 \\ 
   & Comp &  & 0.889 & 0.892 & 0.893 & 0.892 &  & 0.941 & 0.944 & 0.944 & 0.944 &  & 0.986 & 0.987 & 0.987 & 0.988 \\ 
   & HM &  & 0.889 & 0.892 & 0.893 & 0.893 &  & 0.942 & 0.944 & 0.945 & 0.945 &  & 0.986 & 0.987 & 0.988 & 0.988 \\ 
   & S &  & 0.889 & 0.892 & 0.893 & 0.893 &  & 0.942 & 0.944 & 0.945 & 0.945 &  & 0.987 & 0.988 & 0.988 & 0.988 \\ 
   & Naive &  & 0.881 & 0.887 & 0.888 & 0.888 &  & 0.934 & 0.938 & 0.939 & 0.939 &  & 0.979 & 0.981 & 0.981 & 0.982 \\ \hline
  $\exp$& Cal &  & 0.891 & 0.893 & 0.894 & 0.894 &  & 0.943 & 0.944 & 0.945 & 0.945 &  & 0.985 & 0.986 & 0.986 & 0.987 \\ 
   & noBC &  & 0.890 & 0.894 & 0.896 & 0.896 &  & 0.941 & 0.944 & 0.945 & 0.945 &  & 0.985 & 0.986 & 0.986 & 0.986 \\ 
   & Add &  & 0.887 & 0.890 & 0.892 & 0.891 &  & 0.938 & 0.941 & 0.941 & 0.941 &  & 0.983 & 0.983 & 0.984 & 0.984 \\ 
   & Mult &  & 0.888 & 0.892 & 0.893 & 0.893 &  & 0.939 & 0.942 & 0.942 & 0.942 &  & 0.983 & 0.984 & 0.984 & 0.984 \\ 
   & Comp &  & 0.888 & 0.891 & 0.893 & 0.892 &  & 0.939 & 0.941 & 0.942 & 0.942 &  & 0.983 & 0.984 & 0.984 & 0.984 \\ 
   & HM &  & 0.888 & 0.892 & 0.893 & 0.893 &  & 0.939 & 0.942 & 0.942 & 0.943 &  & 0.983 & 0.984 & 0.984 & 0.984 \\ 
   & S &  & 0.888 & 0.893 & 0.904 & 0.917 &  & 0.937 & 0.936 & 0.934 & 0.939 &  & 0.981 & 0.977 & 0.967 & 0.963 \\ 
   & Naive &  & 0.881 & 0.886 & 0.888 & 0.888 &  & 0.934 & 0.938 & 0.939 & 0.939 &  & 0.979 & 0.981 & 0.981 & 0.982 \\ \hline
 $Q_{0.25}$ & Cal &  & 0.891 & 0.894 & 0.894 & 0.895 &  & 0.943 & 0.945 & 0.945 & 0.946 &  & 0.986 & 0.986 & 0.987 & 0.987 \\ 
   & noBC &  & 0.894 & 0.897 & 0.897 & 0.898 &  & 0.945 & 0.947 & 0.948 & 0.948 &  & 0.988 & 0.989 & 0.989 & 0.989 \\ 
   & Add &  & 0.890 & 0.893 & 0.894 & 0.894 &  & 0.942 & 0.944 & 0.944 & 0.944 &  & 0.986 & 0.987 & 0.987 & 0.987 \\ 
   & Mult &  & 0.892 & 0.894 & 0.895 & 0.895 &  & 0.943 & 0.945 & 0.945 & 0.946 &  & 0.986 & 0.987 & 0.987 & 0.987 \\ 
   & Comp &  & 0.891 & 0.894 & 0.895 & 0.895 &  & 0.942 & 0.945 & 0.945 & 0.945 &  & 0.986 & 0.987 & 0.987 & 0.987 \\ 
   & HM &  & 0.892 & 0.894 & 0.895 & 0.895 &  & 0.943 & 0.945 & 0.946 & 0.946 &  & 0.986 & 0.987 & 0.987 & 0.987 \\ 
   & S &  & 0.897 & 0.900 & 0.901 & 0.901 &  & 0.947 & 0.949 & 0.950 & 0.950 &  & 0.988 & 0.989 & 0.989 & 0.989 \\ 
   & Naive &  & 0.884 & 0.888 & 0.890 & 0.890 &  & 0.936 & 0.940 & 0.941 & 0.940 &  & 0.980 & 0.981 & 0.982 & 0.982 \\ \hline
  $Q_{0.75}$ & Cal &  & 0.890 & 0.892 & 0.893 & 0.893 &  & 0.942 & 0.943 & 0.944 & 0.944 &  & 0.985 & 0.986 & 0.986 & 0.986 \\ 
   & noBC &  & 0.890 & 0.893 & 0.894 & 0.894 &  & 0.942 & 0.945 & 0.945 & 0.945 &  & 0.986 & 0.987 & 0.988 & 0.988 \\ 
   & Add &  & 0.887 & 0.889 & 0.890 & 0.890 &  & 0.939 & 0.941 & 0.941 & 0.942 &  & 0.985 & 0.985 & 0.986 & 0.986 \\ 
   & Mult &  & 0.888 & 0.891 & 0.892 & 0.891 &  & 0.940 & 0.942 & 0.943 & 0.943 &  & 0.985 & 0.986 & 0.986 & 0.986 \\ 
   & Comp &  & 0.888 & 0.890 & 0.891 & 0.891 &  & 0.940 & 0.942 & 0.942 & 0.942 &  & 0.985 & 0.986 & 0.986 & 0.986 \\ 
   & HM &  & 0.888 & 0.891 & 0.892 & 0.891 &  & 0.940 & 0.943 & 0.943 & 0.943 &  & 0.985 & 0.986 & 0.986 & 0.986 \\ 
   & S &  & 0.879 & 0.881 & 0.883 & 0.882 &  & 0.934 & 0.936 & 0.937 & 0.937 &  & 0.983 & 0.984 & 0.984 & 0.984 \\ 
   & Naive &  & 0.880 & 0.885 & 0.886 & 0.886 &  & 0.933 & 0.937 & 0.938 & 0.938 &  & 0.978 & 0.980 & 0.981 & 0.981 \\ \hline
 $PG$& Cal &  & 0.893 & 0.896 & 0.906 & 0.914 &  & 0.944 & 0.947 & 0.952 & 0.956 &  & 0.986 & 0.987 & 0.988 & 0.989 \\ 
   & noBC &  & 0.909 & 0.915 & 0.921 & 0.926 &  & 0.957 & 0.957 & 0.958 & 0.960 &  & 0.989 & 0.987 & 0.987 & 0.988 \\ 
   & Add &  & 0.905 & 0.910 & 0.914 & 0.916 &  & 0.953 & 0.953 & 0.953 & 0.954 &  & 0.987 & 0.985 & 0.984 & 0.985 \\ 
   & Mult &  & 0.907 & 0.912 & 0.918 & 0.921 &  & 0.954 & 0.954 & 0.955 & 0.956 &  & 0.988 & 0.986 & 0.986 & 0.986 \\ 
   & Comp &  & 0.906 & 0.911 & 0.918 & 0.923 &  & 0.954 & 0.954 & 0.955 & 0.958 &  & 0.988 & 0.986 & 0.985 & 0.987 \\ 
   & HM &  & 0.907 & 0.912 & 0.918 & 0.923 &  & 0.954 & 0.954 & 0.956 & 0.958 &  & 0.988 & 0.986 & 0.986 & 0.987 \\ 
   & S &  & 0.900 & 0.890 & 0.885 & 0.884 &  & 0.945 & 0.933 & 0.925 & 0.922 &  & 0.984 & 0.977 & 0.971 & 0.966 \\ 
   & Naive &  & 0.885 & 0.891 & 0.905 & 0.914 &  & 0.937 & 0.942 & 0.949 & 0.954 &  & 0.980 & 0.982 & 0.985 & 0.986 \\ \hline
  $Gini$ & Cal &  & 0.909 & 0.909 & 0.909 & 0.909 &  & 0.955 & 0.955 & 0.955 & 0.955 &  & 0.990 & 0.990 & 0.989 & 0.989 \\ 
   & noBC &  & 0.911 & 0.911 & 0.911 & 0.910 &  & 0.957 & 0.957 & 0.957 & 0.957 &  & 0.992 & 0.992 & 0.992 & 0.992 \\ 
   & Add &  & 0.907 & 0.908 & 0.907 & 0.907 &  & 0.954 & 0.954 & 0.954 & 0.954 &  & 0.990 & 0.990 & 0.990 & 0.990 \\ 
   & Mult &  & 0.908 & 0.909 & 0.909 & 0.908 &  & 0.955 & 0.955 & 0.955 & 0.955 &  & 0.991 & 0.991 & 0.991 & 0.991 \\ 
   & Comp &  & 0.908 & 0.909 & 0.908 & 0.908 &  & 0.955 & 0.955 & 0.955 & 0.954 &  & 0.991 & 0.991 & 0.991 & 0.990 \\ 
   & HM &  & 0.909 & 0.909 & 0.909 & 0.908 &  & 0.955 & 0.955 & 0.955 & 0.955 &  & 0.991 & 0.991 & 0.991 & 0.991 \\ 
   & S &  & 0.905 & 0.905 & 0.905 & 0.904 &  & 0.953 & 0.952 & 0.952 & 0.952 &  & 0.990 & 0.990 & 0.990 & 0.990 \\ 
   & Naive &  & 0.890 & 0.890 & 0.890 & 0.889 &  & 0.941 & 0.941 & 0.941 & 0.941 &  & 0.982 & 0.982 & 0.982 & 0.982 \\ 
   \bottomrule\bottomrule
\end{tabular}
}
\end{table}

\begin{table}[H]
\captionsetup{font=footnotesize}
\centering
\caption{\label{nonsamp_CI} The coverage probabilities of confidence intervals for non-sampled areas. Naive and Cal are the naive and the calibrated confidence interval, respectively. The remainder in the Method column is the normal theory confidence interval with the corresponding MSE estimator.}
\scalebox{0.5}{
\begin{tabular}{lllrrrrlrrrrlrrrr}
  \toprule
    \toprule
     \multirow{2}{*}{Parameter} & \multirow{2}{*}{Method} & &\multicolumn{4}{c}{$CP = 0.90$}& & \multicolumn{4}{c}{$CP = 0.95$}& & \multicolumn{4}{c}{$CP = 0.99$} \\
     \cmidrule{4-7} \cmidrule{9-12} \cmidrule{14-17}
    &  & &0.5 & 1 & 2 & 3 && 0.5 & 1 & 2 & 3&& 0.5 & 1 & 2 & 3  \\   \midrule
$\bar{Y}$ & Cal &  & 0.887 & 0.889 & 0.890 & 0.891 &  & 0.942 & 0.942 & 0.943 & 0.943 &  & 0.987 & 0.987 & 0.987 & 0.987 \\ 
   & noBC &  & 0.886 & 0.889 & 0.890 & 0.891 &  & 0.940 & 0.942 & 0.942 & 0.943 &  & 0.988 & 0.990 & 0.990 & 0.990 \\ 
   & Add &  & 0.879 & 0.884 & 0.885 & 0.886 &  & 0.935 & 0.938 & 0.938 & 0.939 &  & 0.985 & 0.987 & 0.987 & 0.987 \\ 
   & Mult &  & 0.881 & 0.886 & 0.886 & 0.888 &  & 0.936 & 0.939 & 0.939 & 0.940 &  & 0.985 & 0.987 & 0.988 & 0.988 \\ 
   & Comp &  & 0.881 & 0.885 & 0.886 & 0.887 &  & 0.936 & 0.939 & 0.939 & 0.940 &  & 0.985 & 0.987 & 0.988 & 0.988 \\ 
   & HM &  & 0.881 & 0.886 & 0.886 & 0.888 &  & 0.936 & 0.939 & 0.939 & 0.940 &  & 0.986 & 0.987 & 0.988 & 0.988 \\ 
   & S &  & 0.898 & 0.902 & 0.903 & 0.903 &  & 0.948 & 0.950 & 0.951 & 0.952 &  & 0.990 & 0.992 & 0.993 & 0.993 \\ 
   & Naive &  & 0.876 & 0.880 & 0.881 & 0.882 &  & 0.930 & 0.933 & 0.933 & 0.934 &  & 0.977 & 0.980 & 0.980 & 0.980 \\ \hline
  $\exp$ & Cal &  & 0.887 & 0.889 & 0.890 & 0.891 &  & 0.942 & 0.942 & 0.943 & 0.943 &  & 0.986 & 0.987 & 0.987 & 0.987 \\ 
   & noBC &  & 0.893 & 0.909 & 0.927 & 0.935 &  & 0.942 & 0.948 & 0.946 & 0.949 &  & 0.985 & 0.978 & 0.969 & 0.968 \\ 
   & Add &  & 0.883 & 0.898 & 0.910 & 0.892 &  & 0.935 & 0.941 & 0.935 & 0.916 &  & 0.981 & 0.975 & 0.962 & 0.943 \\ 
   & Mult &  & 0.885 & 0.902 & 0.920 & 0.925 &  & 0.936 & 0.943 & 0.941 & 0.940 &  & 0.982 & 0.976 & 0.965 & 0.960 \\ 
   & Comp &  & 0.885 & 0.901 & 0.919 & 0.925 &  & 0.936 & 0.942 & 0.940 & 0.940 &  & 0.982 & 0.976 & 0.965 & 0.960 \\ 
   & HM &  & 0.886 & 0.902 & 0.921 & 0.927 &  & 0.937 & 0.943 & 0.941 & 0.942 &  & 0.982 & 0.976 & 0.966 & 0.962 \\ 
   & S &  & 0.900 & 0.913 & 0.927 & 0.933 &  & 0.947 & 0.950 & 0.946 & 0.947 &  & 0.986 & 0.980 & 0.969 & 0.966 \\ 
   & Naive &  & 0.876 & 0.880 & 0.881 & 0.882 &  & 0.930 & 0.933 & 0.933 & 0.934 &  & 0.977 & 0.980 & 0.980 & 0.980 \\ \hline
  $Q_{0.25}$ & Cal &  & 0.887 & 0.889 & 0.890 & 0.891 &  & 0.942 & 0.942 & 0.943 & 0.943 &  & 0.986 & 0.987 & 0.987 & 0.987 \\ 
   & noBC &  & 0.887 & 0.890 & 0.890 & 0.891 &  & 0.941 & 0.942 & 0.943 & 0.943 &  & 0.988 & 0.989 & 0.990 & 0.990 \\ 
   & Add &  & 0.880 & 0.884 & 0.885 & 0.886 &  & 0.935 & 0.938 & 0.938 & 0.939 &  & 0.985 & 0.987 & 0.987 & 0.987 \\ 
   & Mult &  & 0.882 & 0.886 & 0.886 & 0.888 &  & 0.936 & 0.939 & 0.939 & 0.940 &  & 0.985 & 0.987 & 0.988 & 0.988 \\ 
   & Comp &  & 0.881 & 0.885 & 0.886 & 0.887 &  & 0.936 & 0.939 & 0.939 & 0.940 &  & 0.985 & 0.987 & 0.988 & 0.988 \\ 
   & HM &  & 0.882 & 0.886 & 0.886 & 0.888 &  & 0.937 & 0.939 & 0.939 & 0.940 &  & 0.986 & 0.987 & 0.988 & 0.988 \\ 
   & S &  & 0.898 & 0.902 & 0.903 & 0.903 &  & 0.948 & 0.950 & 0.951 & 0.952 &  & 0.990 & 0.992 & 0.993 & 0.993 \\ 
   & Naive &  & 0.877 & 0.880 & 0.881 & 0.882 &  & 0.931 & 0.933 & 0.934 & 0.934 &  & 0.978 & 0.980 & 0.980 & 0.980 \\ \hline
 $Q_{0.75}$ & Cal &  & 0.888 & 0.889 & 0.890 & 0.891 &  & 0.942 & 0.942 & 0.942 & 0.943 &  & 0.986 & 0.987 & 0.987 & 0.987 \\ 
   & noBC &  & 0.887 & 0.889 & 0.890 & 0.891 &  & 0.941 & 0.942 & 0.942 & 0.943 &  & 0.988 & 0.989 & 0.990 & 0.990 \\ 
   & Add &  & 0.880 & 0.884 & 0.885 & 0.886 &  & 0.935 & 0.938 & 0.938 & 0.939 &  & 0.985 & 0.987 & 0.987 & 0.987 \\ 
   & Mult &  & 0.881 & 0.885 & 0.886 & 0.888 &  & 0.936 & 0.939 & 0.939 & 0.940 &  & 0.985 & 0.987 & 0.988 & 0.988 \\ 
   & Comp &  & 0.881 & 0.885 & 0.886 & 0.887 &  & 0.936 & 0.939 & 0.939 & 0.940 &  & 0.985 & 0.987 & 0.988 & 0.988 \\ 
   & HM &  & 0.882 & 0.885 & 0.887 & 0.888 &  & 0.937 & 0.939 & 0.939 & 0.940 &  & 0.986 & 0.987 & 0.988 & 0.988 \\ 
   & S &  & 0.897 & 0.901 & 0.903 & 0.903 &  & 0.948 & 0.950 & 0.951 & 0.952 &  & 0.990 & 0.992 & 0.992 & 0.993 \\ 
   & Naive &  & 0.876 & 0.880 & 0.881 & 0.882 &  & 0.930 & 0.933 & 0.933 & 0.935 &  & 0.978 & 0.980 & 0.980 & 0.980 \\ \hline
  $PG$ & Cal &  & 0.891 & 0.890 & 0.902 & 0.895 &  & 0.944 & 0.941 & 0.946 & 0.944 &  & 0.987 & 0.986 & 0.986 & 0.986 \\ 
   & noBC &  & 0.914 & 0.913 & 0.899 & 0.891 &  & 0.948 & 0.939 & 0.929 & 0.927 &  & 0.978 & 0.971 & 0.970 & 0.979 \\ 
   & Add &  & 0.907 & 0.910 & 0.898 & 0.890 &  & 0.944 & 0.936 & 0.927 & 0.926 &  & 0.976 & 0.969 & 0.968 & 0.976 \\ 
   & Mult &  & 0.909 & 0.912 & 0.899 & 0.891 &  & 0.945 & 0.937 & 0.928 & 0.927 &  & 0.977 & 0.970 & 0.969 & 0.977 \\ 
   & Comp &  & 0.909 & 0.911 & 0.899 & 0.891 &  & 0.945 & 0.937 & 0.928 & 0.926 &  & 0.977 & 0.969 & 0.969 & 0.976 \\ 
   & HM &  & 0.909 & 0.912 & 0.899 & 0.891 &  & 0.945 & 0.937 & 0.928 & 0.927 &  & 0.977 & 0.970 & 0.969 & 0.977 \\ 
   & S &  & 0.923 & 0.921 & 0.908 & 0.900 &  & 0.954 & 0.945 & 0.937 & 0.937 &  & 0.982 & 0.976 & 0.977 & 0.985 \\ 
   & Naive &  & 0.880 & 0.881 & 0.941 & 0.942 &  & 0.933 & 0.933 & 0.967 & 0.968 &  & 0.979 & 0.981 & 0.989 & 0.989 \\ \hline
  $Gini$ & Cal &  & 0.907 & 0.906 & 0.907 & 0.906 &  & 0.954 & 0.953 & 0.954 & 0.953 &  & 0.989 & 0.989 & 0.989 & 0.989 \\ 
   & noBC &  & 0.908 & 0.907 & 0.908 & 0.907 &  & 0.955 & 0.954 & 0.955 & 0.954 &  & 0.991 & 0.991 & 0.991 & 0.991 \\ 
   & Add &  & 0.904 & 0.904 & 0.904 & 0.904 &  & 0.952 & 0.951 & 0.952 & 0.951 &  & 0.990 & 0.989 & 0.990 & 0.989 \\ 
   & Mult &  & 0.906 & 0.905 & 0.906 & 0.905 &  & 0.953 & 0.952 & 0.953 & 0.952 &  & 0.990 & 0.990 & 0.990 & 0.990 \\ 
   & Comp &  & 0.905 & 0.905 & 0.905 & 0.905 &  & 0.953 & 0.952 & 0.953 & 0.952 &  & 0.990 & 0.990 & 0.990 & 0.990 \\ 
   & HM &  & 0.906 & 0.905 & 0.906 & 0.905 &  & 0.953 & 0.952 & 0.953 & 0.952 &  & 0.990 & 0.990 & 0.990 & 0.990 \\ 
   & S &  & 0.907 & 0.906 & 0.907 & 0.907 &  & 0.954 & 0.953 & 0.954 & 0.953 &  & 0.990 & 0.990 & 0.990 & 0.990 \\ 
   & Naive &  & 0.886 & 0.885 & 0.886 & 0.886 &  & 0.938 & 0.937 & 0.938 & 0.937 &  & 0.981 & 0.981 & 0.981 & 0.980 \\ 
   \bottomrule\bottomrule
\end{tabular}
}
\end{table}

\section{Implementation of Jackknife Variance Estimator for Model Parameter Estimators}

Let the model for estimation be the nested error linear regression model given by 
\begin{align}\label{sampledist} 
    y_{ij} = \beta_{0} + \bm{x}_{ij}'\bm{\beta}_{1} + u_{i } + e_{ij}, j \in s_{i}, i\in s,
\end{align}
where $u_{i} \stackrel{iid}{\sim} N(0,\sigma^{2}_{u})$ and $e_{ij} \stackrel{iid}{\sim} N(0, \sigma^{2}_{e})$. Assume that the sampling weight $w_{ij}$ within the selected areas satisfies 
\begin{align}\label{weightmodel} 
    E_{si}[w_{ij} \mid \bm{x}_{ij}, y_{ij}, u_{i} , I_{i} = 1] = \kappa_{i}\mbox{exp}( \bm{x}_{ij}'\bm{\gamma}_{1} + \gamma_{2} y_{ij} + \bm{x}_{ij}'\bm{\gamma}_{3} y_{ij}), 
\end{align}
where $\kappa_{i} = N_{i}^{-1}\sum_{j = 1}^{N_{i}}\mbox{exp}( -\bm{x}_{ij}'\bm{\gamma}_{1} - \gamma_{2} y_{ij} - \bm{x}_{ij}'\bm{\gamma}_{3} y_{ij})$. To construct predictors for nonsampled areas, we postulate a further assumption that the area-level weight $w_{i}$ satisfies a lognormal model given by 
\begin{align*}
    \mbox{log}(w_{i}) \mid u_{i}, I_{i} = 1 \sim N(\lambda_{1} + \lambda_{2} u_{i}, \tau^{2}),
\end{align*}
such that $E_{s}[w_{i} \mid u_{i}] = \mbox{exp}(\lambda_{1} + \lambda_{2} u_{i} + \tau^{2}/2)$.  Decompose the vector $\bm{\psi}$ as $\bm{\psi} = (\bm{\psi}_{s}', \bm{\psi}_{ns}')'$, where $\bm{\psi}_{s} = (\bm{\beta}', \sigma^{2}_{u}, \sigma^{2}_{e}, \gamma_{2}, \bm{\gamma}_{3}')'$ and $\bm{\psi}_{ns} = (\lambda_{1}, \lambda_{2}, \tau^{2})'$. Likewise, express the estimator $\hat{\bm{\psi}}$ as  $\hat{\bm{\psi}} = (\hat{\bm{\psi}}_{s}', \hat{\bm{\psi}}_{ns}')'$, where $\hat{\bm{\psi}}_{s} = (\hat{\bm{\beta}}', \hat{\sigma}^{2}_{u}, \hat{\sigma}^{2}_{e}, \hat{\gamma}_{2}, \hat{\bm{\gamma}}_{3}')'$ and $\hat{\bm{\psi}}_{ns} = (\hat{\lambda}_{1}, \hat{\lambda}_{2}, \hat{\tau}^{2})'$. We define a jackknife estimator of the variance of $\hat{\bm{\psi}}$. To motivate the jackknife estimator, observed that $\hat{\bm{\psi}}_{s}$ for the linear model can be expressed as  $\hat{\bm{\psi}}_{s} = argmax_{\bm{\psi}_{s}}m(\bm{\psi}_{s})$, where
\begin{align}\label{objfunmax} 
m(\bm{\psi}_{s}) & = \sum_{i = 1}^{d}-\frac{1}{2}\left\{ \mbox{log}(|\bm{V}_{i}|) + (\bm{y}_{si} - \bm{X}_{i}\bm{\beta})\bm{V}_{i}^{-1}(\bm{y}_{si} - \bm{X}_{i}\bm{\beta})\right\} \nonumber \\ & - \sum_{i = 1}^{d}\sum_{j \in s_{i}} (w_{ij} - \kappa_{i}\mbox{exp}( \bm{x}_{ij}'\bm{\gamma}_{1} + \gamma_{2} y_{ij} + \bm{x}_{ij}'\bm{\gamma}_{3}y_{ij}))^{2}, 
\end{align}
$\bm{V}_{i} = \sigma^{2}_{u}\bm{1}_{n_{i}}\bm{1}_{n_{i}}' + \sigma^{2}_{e}\bm{I}_{n_{i}\times n_{i}}$ for $i \in s$ and $\bm{X}_{i} = (\bm{x}_{i1},\ldots, \bm{x}_{i n_{i}})'$. Define $$\hat{\bm{\psi}}_{s-(i)} = (\hat{\bm{\beta}}_{-(i)}', \hat{\sigma}^{2}_{u-(i)}, \hat{\sigma}^{2}_{e-(i)}, \hat{\gamma}_{2-(i)}, \hat{\bm{\gamma}}_{3-(i)}')'$$ as $\hat{\bm{\psi}}_{s-(i)} = argmax_{\bm{\psi}_{s}}m_{-(i)}(\bm{\psi}_{s})$, where
\begin{align}\label{objfunmaxminusi} 
m_{-(i)}(\bm{\psi}_{s}) & = \sum_{k\in s, k\neq i} -\frac{1}{2}\left\{ \mbox{log}(|\bm{V}_{k}|) + (\bm{y}_{sk} - \bm{X}_{k}\bm{\beta})\bm{V}_{k}^{-1}(\bm{y}_{sk} - \bm{X}_{k}\bm{\beta})\right\}  \\ & - \sum_{k \in s, k\neq i} \sum_{j \in s_{k}} (w_{kj} - \kappa_{k}\mbox{exp}( \bm{x}_{kj}'\bm{\gamma}_{1} + \gamma_{2} y_{kj} + \bm{x}_{kj}'\bm{\gamma}_{3}y_{kj}))^{2}.
\end{align}
Define 
\begin{align}\label{psinsminusi} 
    \hat{\bm{\psi}}_{ns-(i)} 
    &= argmax_{\Theta } \prod_{k\in s, k\neq i} \int_{-\infty}^{\infty}\frac{1}{\tau} \phi\left( \frac{ \mbox{log}(w_{k}) - \lambda_{1} - \lambda_{2}  u_{k} }{  \tau}\right)\hat{f}_{s}(u_{k} \mid D_{s}) du_{k}\\
    &=argmax_{\Theta } \prod_{k\in s, k\neq i} \frac{1}{\sqrt{\lambda_{2}^{2}}}\frac{1}{\sqrt{2\pi\big(\frac{\tau^{2}}{\lambda_{2}^{2}}+\hat{v}_{k}^{2}\big)}}exp\bigg(-\frac{\big(\frac{log(w_{k})-\lambda_{1}}{\lambda_{2}}-\hat{u}_{k})^{2}}{2\big(\frac{\tau^{2}}{\lambda_{2}^{2}}+\hat{v}_{k}^{2}\big)}\bigg)
\end{align}
where $\Theta = (-\infty, \infty) \times (-\infty, \infty) \times (0,\infty)$, and $\hat{f}_{s}(u_{k} \mid D_{s})$ is the density of a normal distribution with mean $\hat{u}_{k}(\hat{\bm{\beta}}, \hat{\sigma}^{2}_{u}, \hat{\sigma}_{e}^{2}) = \hat{\sigma}^{2}_{u}(\hat{\sigma}^{2}_{u} + \hat{\sigma}^{2}_{e}/n_{k})^{-1}(\bar{y}_{k} - (1, \bar{\bm{x}}_{k}')\hat{\bm{\beta}})  $ and variance $\hat{\sigma}_{u}^{2}\hat{\sigma}^{2}_{e}n_{k}^{-1}(\hat{\sigma}^{2}_{u} + \hat{\sigma}^{2}_{e}/n_{k})^{-1}$. Note that when calculating $\hat{\bm{\psi}}_{ns-(i)}$, $\hat{f}_{s}(u_{k} \mid D_{s})$ is considered as given, which implies that we do not reflect the variability from $\hat{\bm{\psi}}_{s}$. Thus, we define an estimator of the variance of $\hat{\bm{\psi}}$ by $\hat{\bm{V}}_{J} = \mbox{block-diag}(\hat{\bm{V}}_{J,s}, \hat{\bm{V}}_{J,ns})$, where 
\begin{align*}
    \hat{\bm{V}}_{J,s}  & = \frac{d -1}{d}\sum_{i=1}^{d}(\hat{\bm{\psi}}_{s-(i)} - \bar{\bm{\psi}}_{s})(\hat{\bm{\psi}}_{s-(i)} - \bar{\bm{\psi}}_{s})', \\ \nonumber 
    \hat{\bm{V}}_{J,ns} & = \frac{d -1}{d}\sum_{i=1}^{d}(\hat{\bm{\psi}}_{ns-(i)} - \bar{\bm{\psi}}_{ns})(\hat{\bm{\psi}}_{ns-(i)} - \bar{\bm{\psi}}_{ns})',
\end{align*}
$\bar{\bm{\psi}}_{s} = d^{-1}\sum_{i=1}^{d} \hat{\bm{\psi}}_{s-(i)} $, and $\bar{\bm{\psi}}_{ns} = d^{-1}\sum_{i=1}^{d} \hat{\bm{\psi}}_{ns-(i)}$.  Assume that 
\begin{align}\label{vjnormal} 
    \hat{\bm{V}}_{J}^{-0.5}(\hat{\bm{\psi}} - \bm{\psi}) \stackrel{L}{\rightarrow} N(\bm{0}, \bm{I}_{q}),
\end{align}
where $q$ is the dimension of $\bm{\psi}$.  We then define an estimator of $M_{2i}$ as 
\begin{align}\label{Mhat2iR} 
    \hat{M}_{2i,R} = B^{-1}\sum_{b = 1}^{B}(\hat{\theta}_{i,R}(\hat{\bm{\psi}}^{(b)}) - \hat{\theta}_{i,R}(\hat{\bm{\psi}}))^{2}, 
\end{align}
where $\hat{\bm{\psi}}^{(b)} \stackrel{iid}{\sim} N(\hat{\bm{\psi}}, \hat{\bm{V}}_{J})$. Negative values for the variance estimates could be generated from $N(\hat{\bm{\psi}},\hat{V}_{J})$, especially for small $d$.  To avoid negative variance estimates, we first use the delta method to obtain the distribution of the log-transformed variance estimates. We then simulate estimates on the log scale and subsequently apply a back-transformation. 

Define the log-transformed variance estimators in $\hat{\bm{\psi}}$  as 
\begin{align*}
  \bm{g}(\hat{\bm{\psi}}) = (\hat{\bm{\beta}}', log(\hat{\sigma}^{2}_{u}), log(\hat{\sigma}^{2}_{e}),  \hat{\gamma}_{2} , \hat{\bm{\gamma}}_{3}^{T},\hat{\lambda}_{1},\hat{\lambda}_{2},log(\hat{\tau}^{2}))'.  
\end{align*}
Under equation (\ref{vjnormal}), using the delta method, the distribution of $\bm{g}(\hat{\bm{\psi}})$ can be approximated as $N\big(\bm{g}(\hat{\bm{\psi}}), \hat{J}_{g}\hat{\bm{V}}_{J}\hat{J}_{g}^{T}\big),$ where $\hat{J}_{g}$ is the jacobian of $\bm{g}$ defined as
\begin{align}
    \hat{\mathbb{J}}_{g} =\frac{d\bm{g}(\bm{\psi})}{d\bm{\psi}} \biggl\lvert_{\bm{\psi} = \hat{\bm{\psi}}} 
    ={\rm diag}(\bm{1}_{p}^{T}, 1/\hat{\sigma}^{2}_{u}, 1/\hat{\sigma}^{2}_{e},  1 ,\bm{1}_{r}^{T},1, 1,1/\hat{\tau}^{2}),\nonumber
\end{align}
$p$ is the dimension of $\bm{\beta}$ and $r$ is the dimension of $\bm{\gamma}_{3}$. Then, we can address the support problem by using $\bm{g}^{-1}(\bm{g}(\hat{\bm{\psi}})^{(b)})$, where $\hat{\bm{\psi}}^{(b)} = \bm{g}(\hat{\bm{\psi}})^{(b)}\stackrel{iid}{\sim}N\big(\bm{g}(\hat{\bm{\psi}}), \hat{J}_{g}\hat{\bm{V}}_{J}\hat{J}_{g}^{T}\big)$ and $\bm{g}^{-1}(\bm{\psi}) = (\bm{\beta}',\exp(\sigma_{u}^{2}),\exp(\sigma_{e}^{2}),\gamma_{2},\bm{\gamma}_{3}^{T},\lambda_{1},\lambda_{2},\exp(\tau^{2}))'$.

\end{document}